\newif\ifUsenix\Usenixfalse %

\ifUsenix
    \documentclass[letterpaper,twocolumn,10pt]{article}
    \usepackage{usenix2019_v3}
    \usepackage{authblk}
\else
     \documentclass[conference,compsoc]{IEEEtran}
     \usepackage{xurl}
     \usepackage[noadjust]{cite}

\fi

\newcommand{\isAnon}{0}

\usepackage[dvipsnames]{xcolor}

\definecolor{uzlmain}{RGB}{0,75,90}
\definecolor{uzlpal6}{RGB}{198,220,226}

\usepackage{ifthen}
\usepackage{booktabs}
\usepackage{listings}
\usepackage{amsmath}
\usepackage{amssymb}
\usepackage{mathtools}
\usepackage{physics}

\usepackage{caption}
\usepackage{subcaption}

\usepackage{multirow}

\usepackage[normalem]{ulem} %
\usepackage{tcolorbox} %
\tcbset{colback=uzlpal6!50,colframe=uzlmain} %

\usepackage{fontawesome} %

\usepackage{siunitx}
\usepackage{tikz}

\usepackage[]{hyperref}         %
\hypersetup{                    %
  colorlinks,
  linkcolor={green!80!black},
  citecolor={red!70!black},
  urlcolor={blue!70!black}
}

\usepackage[capitalise]{cleveref}

\usepackage{orcidlink}

\usepackage{float}

\usepackage{pifont}
\newcommand{\greencheck}{{\color{OliveGreen}{\ding{52}}}}
\newcommand{\redx}{{\color{Mahogany}{\ding{56}}}}
\newcommand{\blueneutral}{{\color{MidnightBlue}{\ding{71}}}}

\usepackage{todonotes}

\renewcommand{\texttt}[1]{$\mathtt{#1}$}

\newcommand{\tool}{\textsc{Obelix}}

\newcommand{\toolI}{\textsc{I-FixedLength}}
\newcommand{\toolII}{\textsc{II-FixedCount}}
\newcommand{\toolIII}{\textsc{III-FixedPattern}}
\newcommand{\toolIV}{\textsc{IV-AlignedPattern}}
\newcommand{\toolV}{\textsc{V-Ciphertext}}

\newcommand{\toolIshort}{\textsc{I-FixedLength}} %
\newcommand{\toolIIshort}{\textsc{II-FixedCount}} %
\newcommand{\toolIIIshort}{\textsc{III-FixedPat.}} %
\newcommand{\toolIVshort}{\textsc{IV-AlignedPat.}} %
\newcommand{\toolVshort}{\textsc{ V-Ciphertext}} %

\newcommand{\timingleakage}{\faClockO{}}
\newcommand{\singlestepleakage}{\faPaw{}}
\newcommand{\ciphertextleakage}{\faStarHalfO}

\newcommand{\timebase}{\shortstack{orig\\ ($\mu\mathrm{s}$)}}
\newcommand{\timetool}{time}
\newcommand{\overhead}{factor}

\makeatletter
\def\@IEEEsectpunct{.\ \,}
\def\paragraph{\@startsection{paragraph}{4}{\z@}{1.5ex plus 1.5ex minus 0.5ex}%
{0ex}{\normalfont\normalsize\itshape}}
\makeatother

\begin{document}

\title{\tool{}: Mitigating Side-Channels Through Dynamic Obfuscation}

\ifthenelse{\equal{\isAnon}{1}}{
	\author{Anonymous Submission}
}{
    \ifUsenix    
    \author[]{Jan Wichelmann}
    \author[]{Anja Rabich}
    \author[]{Anna Pätschke}
    \author[]{Thomas Eisenbarth}
    \affil[]{University of L\"ubeck, L\"ubeck, Germany}
    \affil[]{\textit{\{j.wichelmann,\,a.rabich,\,a.paetschke,\,thomas.eisenbarth\}@uni-luebeck.de}}
  \else

    \author{
        \IEEEauthorblockN{
            Jan Wichelmann~\orcidlink{0000-0002-5748-5462},
            Anja Rabich~\orcidlink{0009-0006-5696-1355},
            Anna Pätschke~\orcidlink{0000-0001-7828-2333},
            Thomas Eisenbarth~\orcidlink{0000-0003-1116-6973} 
        }
        \IEEEauthorblockA{
            Universität zu L\"ubeck, L\"ubeck, Germany\\
            \{a.paetschke,\,j.wichelmann,\,thomas.eisenbarth\}@uni-luebeck.de
        }
    }
  \fi
}

\maketitle

\begin{abstract}

Trusted execution environments (TEEs) offer hardware-assisted means to protect code and data.
However, as shown in numerous results over the years, attackers can use side-channels to leak data access patterns and even single-step the code.
While the vendors are slowly introducing hardware-based countermeasures for some attacks, others will stay unaddressed. This makes a software-level countermeasure desirable, but current available solutions only address very specific attack vectors or have a narrow leakage model.

In this work, we take a holistic view at the vulnerabilities of TEEs and design a tool named \tool{}, which is the first to protect both code and data against a wide range of TEE attacks, from cache attacks over single-stepping to ciphertext side-channels.
We analyze the practically achievable precision of state-of-the-art single-stepping tools, and present an algorithm which uses that knowledge to divide a program into uniform code blocks, that are indistinguishable for a strong attacker.
By storing these blocks and the program data in oblivious RAM, the attacker cannot follow execution, effectively protecting both secret code and data.
We describe how we automate our approach to make it available for developers who are unfamiliar with side-channels.
As an obfuscation tool, \tool{} comes with a considerable performance overhead, but compensates this with strong security guarantees and easy applicability without requiring any expert knowledge.

\end{abstract}

\section{Introduction}
\label{sec:intro}

With ongoing digitization, there is a growing demand for hardware-enforced security that cannot be bypassed by privileged malicious parties on the same system. This applies to moving sensitive data processing to the cloud, but also to local applications which verify the user's identity or enforce the validity of software or content licenses. Owners of sensitive workloads may want to protect both their secret data and their code. Processor vendors recognized these demands and designed a variety of so-called \emph{Trusted Execution Environments} (TEEs) that use hardware access control and cryptography to restrict access to code and data to its owner. 
Examples are Intel SGX~\cite{intel2016sgxguide} for user-space processes, and AMD SEV~\cite{amd2020sev} and Intel TDX~\cite{intel-tdx-manual} for whole-VM protection.

However, in recent years, there have been numerous examples of how gaps in the threat model and implementation issues can be exploited to tamper with the executed code or extract secret data. 
For example, all currently available TEEs are known to be vulnerable to microarchitectural side-channels like TLB and cache attacks~\cite{DBLP:conf/uss/GrasRBG18,bernstein2005cache,DBLP:conf/ctrsa/OsvikST06}. For simplicity, we refer to those as \emph{timing side-channels}. %
An attack method that is more specific to TEEs is \emph{single-stepping}~\cite{DBLP:conf/sosp/BulckPS17,sev-step}, which allows an attacker to precisely measure the execution of single instructions. While those attacks do not allow direct access to the protected data, the attacker gets valuable information about the TEE's execution state, which may be used to partially derive the currently processed data and code, e.g., through counting instructions~\cite{DBLP:conf/uss/MoghimiBHPS20} or measuring their execution time~\cite{DBLP:conf/ccs/BulckPS18,DBLP:conf/uss/PudduSHC21}.
Another TEE-related attack vector are \emph{ciphertext side-channels}~\cite{DBLP:conf/uss/LiZWLC21,DBLP:conf/sp/LiWW0TZ22}, where the attacker frequently reads encrypted memory and observes whether the ciphertext at a specific address changes, breaking common constant-time primitives.

For each of those attacks, there are numerous proposed countermeasures and tools, which focus on protecting the secret data~\cite{DBLP:conf/ccs/GeimerVRDBM23}.
Only few hardware-based methods were deployed by CPU vendors yet, so software-level countermeasures are necessary.
For example, timing attacks can be mitigated by writing constant-time code or linearizing existing code~\cite{DBLP:conf/ccs/BorrelloDQG21}.
Recently, with AEX-Notify~\cite{DBLP:conf/uss/ConstableBCXXAK23}, a hardware-assisted single-stepping countermeasure for Intel SGX was introduced, but that does not apply to other TEEs in the field.
Some libraries employ custom-crafted constant-time code to defend against timing and single-stepping attacks, but that code still remains vulnerable to ciphertext side-channel attacks~\cite{DBLP:conf/sp/LiWW0TZ22}. At the time of writing, the only available automated countermeasure for ciphertext side-channels is Cipherfix~\cite{DBLP:conf/uss/WichelmannPW023}, which relies on dynamic taint tracking and binary rewriting and thus lacks the stability needed for practical deployment.
Writing robust code that is immune to all known attack vectors requires expert knowledge, takes a lot of resources and still is not guaranteed to be side-channel free. Finally, it is not straightforwardly possible to protect the code \emph{itself} against extraction by the attacker, which may be desirable if it, for example, contains secret algorithms.

In this work, we take a broad view at security issues on TEEs. We classify currently known attack vectors against commonly available TEEs, and discuss the suitability of existing software-level defenses, finding that each only addresses a subset of attacks. Subsequently, we show how those attack vectors can be averted with a \emph{single}, modular drop-in countermeasure that does only require minimal action from the user. Contrary to other mitigation approaches, our approach named \tool{} does not only aim to protect secret data, but also the executed algorithms, against all relevant side-channel vulnerabilities.
\tool{} takes and advances the ideas from Obfuscuro~\cite{DBLP:conf/ndss/AhmadJXZSL19}, to build a dynamic obfuscation engine which runs within any TEE and can reliably protect code and data against extraction.
Based upon an evaluation of the \emph{practical} capabilities of a single-stepping attacker, we design an algorithm to partition machine code into a set of uniform code blocks, which are indistinguishable for said attacker.
By storing the code blocks and the associated data in an ORAM, we effectively prevent the attacker from learning anything about the executed code and data.
We present a proof-of-concept implementation for \tool{} as an LLVM compiler extension, and apply it to a number of programs.

\subsection{Our Contribution}
We present \tool{}, a drop-in obfuscation engine that automatically mitigates a wide range of TEE-related attacks.
For that, we:
\begin{itemize}
    \item discuss existing countermeasures and determine the properties needed for a generic mitigation;
    \item evaluate Linear ORAM and Path ORAM for their suitability within an untrusted client setting with small amounts of data;
    \item analyze an attacker's ability to distinguish instructions through single-stepping;
    \item design an algorithm to split a program into uniform code blocks which are indistinguishable to an attacker;
    \item show how to combine these components into a single comprehensive countermeasure.
\end{itemize}

The source code of our proof-of-concept implementation is available at \url{https://github.com/UzL-ITS/obelix}.

\section{Background}
\label{sec:background}

\subsection{Trusted Execution Environments}
\label{sec:TEEs}
Trusted Execution Environments (TEEs) offer hardware-based isolation for protecting a program's execution against privileged adversaries on the same system.
Starting with small user-space enclaves in Intel SGX~\cite{DBLP:journals/iacr/CostanD16} with at most 96MB of storage, modern TEEs are nowadays able to protect whole virtual machines (VMs). %
Common features of TEEs are memory encryption, hardware-enforced memory access prevention and runtime attestation.
However, their implementation differs greatly.
For example, Intel SGX reserves a small amount of physical memory for the enclave, which is integrity protected, features fresh ciphertexts (only SGX version 1~\cite{intel-sgx2-freshness}) and cannot be accessed by any other party than the running enclave.
In contrast, the current version of AMD SEV~\cite{amd2020sev} only actively prevents unauthorized write accesses to enclave memory, while relying on its (deterministic) memory encryption for averting illegitimate reads.

\subsection{Side-Channel Attacks}
TEEs are designed to protect against architectural attackers, i.e., malicious privileged processes on the same system which try to directly tamper with the execution.
However, the threat models of many TEEs exclude \emph{side-channel attacks}~\cite{intel2016sgxguide}.
A side-channel attacker does not access private data directly, but tries to derive the data by observing seemingly unrelated channels. A simple example is measuring the execution time of the program to learn about the length of the processed data.
Due to many performance optimizations and shared resources, the microarchitecture of CPUs exhibits lots of side-channel leakages.
A prominent example are cache attacks, where the attacker learns whether victim data was recently accessed by manipulating and observing the cache state%
~\cite{bernstein2005cache,DBLP:conf/ctrsa/OsvikST06,aciiccmez2007power}.
Other examples are attacks through the TLB~\cite{DBLP:conf/uss/GrasRBG18} or contention of execution ports~\cite{DBLP:conf/sp/AldayaBHGT19}. For simplicity, we refer to those as \emph{timing attacks}.
Software vulnerability to side-channel attacks is mostly rooted in its handling of secret data. If a program makes a secret-dependent memory access or executes a secret-dependent chunk of code, the attacker can use one of the above side-channel attack primitives to learn which data or code was accessed and thus reconstruct the secret.

\subsection{TEE-Specific Side-Channel Attacks}
While there are already numerous side-channel vulnerabilities on common systems, TEEs with their unique threat model are particularly susceptible.
Various attacks~\cite{DBLP:journals/tches/DallMEGHMY18,DBLP:conf/ches/MoghimiIE17,DBLP:conf/woot/BrasserMDKCS17,DBLP:conf/eurosec/GotzfriedESM17,DBLP:conf/dimva/SchwarzWGMM17} show how leakage can be used recover secrets from executions deemed protected.
Not only the parts of the execution inside the TEE but also control mechanisms from outside the protected memory region, like page management, can leak information to the attacker.
SGX uses protected areas for the enclave page tables, but these can still be stealthily monitored via the cache~\cite{DBLP:conf/uss/BulckWKPS17,DBLP:conf/ccs/WangCPZWBTG17}, or the access rights can be manipulated~\cite{DBLP:conf/sp/XuCP15} by the OS.
However, an attacker can not only leak secrets by observing microarchitectural effects of unprotected implementations.
The encrypted memory of VM TEEs is not completely protected from the hypervisor, and the strong attacker scenario allows for very precise and fine-grained observations that enable further attacks.

\subsubsection{Ciphertext side-channel attacks}
The ciphertext side-channel~\cite{DBLP:conf/uss/LiZWLC21,DBLP:conf/sp/LiWW0TZ22} is an example of a more structural leakage.
Ciphertext side-channels are unique to systems that use deterministic memory encryption in combination with read access for a hypervisor to the encrypted memory.
While TEEs like Intel SGX version 1 only protect a small amount of memory and can thus keep fresh nonces needed for non-deterministic encryption, this does not scale for TEEs which manage gigabytes of memory, like VMs.
Thus, TEEs like Intel SGX version 2~\cite{intel-sgx2-freshness} and AMD SEV employ deterministic memory encryption, where the same plaintext written to the same address results in the same ciphertext. However, while Intel SGX prohibits unauthorized reads, AMD SEV allows the malicious hypervisor full read access to the ciphertexts.
If an implementation now does multiple write accesses to the same address depending on a few secret bits, the attacker can label the few resulting (and repeating) ciphertexts and infer the secret~\cite{DBLP:conf/uss/LiZWLC21,DBLP:conf/sp/LiWW0TZ22}.
This attack is unique in a way, as constant-time implementations, which are usually considered the standard countermeasure for many side-channel attacks, are particularly vulnerable.
There are no publicly stated plans of AMD for introducing an effective prevention of ciphertext reads, and it is unclear whether other upcoming TEEs like Intel TDX~\cite{intel-tdx-manual} and ARM CCA~\cite{DBLP:conf/osdi/LiLDGNSS22} are immune to this class of attacks.
Thus, protecting against ciphertext side-channels is left to software.

\subsubsection{Single-stepping attacks}

Another TEE-specific attack vector is \emph{single-stepping}.
As the TEE threat model explicitly considers privileged attackers, they can configure the CPU to precisely interrupt the enclave after every instruction.
With dedicated frameworks such as SGX-Step~\cite{DBLP:conf/sosp/BulckPS17} or SEV-Step~\cite{sev-step}, they can conduct very fine-grained measurements.
For example, by measuring the latency of instructions, it is possible to tell apart certain opcodes and alignments~\cite{DBLP:conf/ccs/BulckPS18,DBLP:conf/uss/PudduSHC21}.
Another powerful primitive is instruction counting to learn loop iteration counts~\cite{DBLP:conf/ccs/BulckOMAGP19} or the length of small intra-cache line secret-dependent branches~\cite{DBLP:conf/uss/MoghimiBHPS20}.
Finally, single-stepping allows for precisely applying cache attacks against the data accessed by a single instruction, enabling exploits of small vulnerabilities with as few as one measurement~\cite{DBLP:conf/ccs/SieckBW021,sev-step}, which is impossible with ordinary cache attacks due to their comparatively low temporal resolution.

\subsection{ORAM}
To defend against timing attackers in the cloud, one needs to conceal the memory access patterns. One such solution is Oblivious RAM (ORAM)~\cite{DBLP:journals/jacm/GoldreichO96}. ORAM was originally designed for large databases which do not fit in a trusted local client and thus are stored on an untrusted server. However, ORAM can also be adapted to aid side-channel defenses for both hiding code and data~\cite{DBLP:conf/uss/RaneLT15,DBLP:conf/ndss/AhmadJXZSL19,DBLP:conf/ndss/SasyGF18}. A straightforward ORAM technique is Linear ORAM, where the client sequentially accesses all data blocks, discarding all but one. While easy to implement, this approach doesn't scale for large amounts of data. Thus, numerous other algorithms were proposed.
One of particular interest for side-channel defense is Path ORAM~\cite{DBLP:journals/jacm/StefanovDSCFRYD18}, where the data is stored in a tree.
On each access, the \emph{path} containing the searched data is sent to the client.
The client then replaces the address of the searched data by a random one, and writes the entire path back into the tree.
Data which could not be written back (because all nodes in the path are already used) is kept in a local \emph{stash} and only written back at the next access.
The mapping of addresses to leafs in the tree is stored in a local \emph{position map}, which may be implemented as a Path ORAM as well, leading to recursively nested Path ORAMs. This way, the protocol offers a logarithmic complexity.

\section{Defenses Against Attacks on TEEs}
\label{sec:defenses}

In the following, we describe the attacker capabilities in a TEE setting and identify gaps in existing hardware-assisted and software-level defense mechanisms.
In line with this, we show that an existing software defense, Obfuscuro~\cite{DBLP:conf/ndss/AhmadJXZSL19}, can be broken through modern single-stepping techniques.

\subsection{Attacker Model}
In line with the standard TEE attacker model, we assume a privileged attacker, who acts as a malicious hypervisor with read access to enclave memory. The memory itself is encrypted. %
The attacker has no way of decrypting the memory or modifying it, and execution state such as register values is considered safe.
The TEE is susceptible to single-stepping attacks, i.e., the attacker can deterministically interrupt the enclave with precise instruction granularity.
This single-stepping capability allows for counting the instructions that are executed in the enclave as well as precise measurements of their individual cache usage, execution latency and, in some cases (integer division), even operand-dependent latency.
We evaluate the practically feasible attacker capabilities in this regard for two exemplary systems in \Cref{sec:evaluation}.
Attack vectors requiring transient execution are considered out of scope.
Adversaries attempting to leverage attacks from within a TEE on the hypervisor or neighboring TEEs are not within the attacker model. Finally, we assume that the code running in the enclave is bug-free and does not exhibit software-level vulnerabilities.

\subsubsection{Goals of the Attacker}
We consider two major adversarial objectives: 
First, they are interested in the (secret) data the program processes. %
Second, they want to learn which algorithms are executed in the enclave, and, by extension, maybe even extracting parts of the executed code. In fact, the logic running in the enclave may be intellectual property that should be kept confidential, or the owner may not want to expose implementation details.
The attacker tries to collect execution traces through various side-channels and reverse engineer the program's control and data flow, aiming to deduce which kinds of algorithms are deployed, and perhaps even identify specific modules.

\subsection{Hardware-Assisted Single-Stepping Prevention}
Most side-channel attacks are considered out-of-scope by TEE vendors and responsibility is shifted to the user (e.g., Intel SGX~\cite[Sec. Protection from Side-Channel Attacks]{intel2016sgxguide}). However, for single-stepping, a few countermeasures are under development.
AEX-Notify~\cite{DBLP:conf/uss/ConstableBCXXAK23} makes Intel SGX enclaves interrupt-aware and can thereby hinder instruction-granular observations.
However, AEX-Notify does not get applied on all older processors~\cite{intel-aex-notify}, and has not yet been thoroughly scrutinized by the research community.
During the evolution of the newer virtual machine TEEs, single-stepping protection gets partially factored in the design process.
For Intel TDX, single-stepping is restricted by randomly limiting the number of instructions executed after an interrupt through the TDX module~\cite[Sec.~17]{intel-tdx-manual}.
At the time of writing, no countermeasures are proposed against single-stepping for AMD SEV and ARM CCA.

\subsection{Software Leakage Defenses}
Due to limited availability of hardware-based security mechanisms, software-level approaches are necessary. For the three main attack classes, timing side-channels, single-stepping and ciphertext side-channels, numerous manual and (semi-)automated countermeasures were proposed. In the following, we broadly classify them and show how they can protect against algorithm fingerprinting and exfiltration of secret data.
The results are summarized in \Cref{table:defenses}.

\begin{table}[t]
    \centering 
    \caption{Defenses against algorithm inference (Alg. inf.) and secret data exfiltration (Sec. data exf.) through the attack classes timing leakage~\timingleakage{}, single-stepping~\singlestepleakage{} and ciphertext side-channels~\ciphertextleakage{}. Checkmarks~\greencheck{} show protection against the given leakage class, crosses~\redx{} that no protection is achieved, and a diamond~\blueneutral{} indicates that protection depends on the granularity of the defense. \tool{} combines oblivious code and data patterns with ciphertext freshness to protect against both algorithm inference and secret data exfiltration.}
    \label{table:defenses}
    \small
    \begin{tabular}{l c c c c c}%
        \toprule
         \multirow{2}[3]{*}{Defense} & \multicolumn{2}{c}{Alg. inf.} & \multicolumn{3}{c}{Sec. data exf.} \\ \cmidrule(lr){2-3} \cmidrule(lr){4-6}
         & \timingleakage{} & \singlestepleakage{} & \timingleakage{} & \singlestepleakage{} & \ciphertextleakage{} \\
        \midrule
        Constant code and data patterns & \redx{} & \redx{} & \greencheck{} & \greencheck{} & \redx{} \\
        Oblivious code and data patterns & \greencheck{} & \blueneutral{} & \greencheck{} & \blueneutral{} & \redx{} \\
        Ciphertext freshness & \redx{} & \redx{} & \redx{} & \redx{} & \greencheck{} \\
        \bottomrule
    \end{tabular}
\end{table}

\subsubsection{Constant code and data patterns}
Constant control flow and constant data access patterns (also called constant-time code) are popular building blocks for secret-independent code.
While the executed code is assumed to be publicly known, the secret data being processed must be protected.
The idea of constant-time code is that for any pair of different secret inputs the program executes the same instructions and accesses the same memory addresses.
An attacker collecting an execution trace via timing side-channels like cache leakage thus should not be able to use the trace to derive information about the secret inputs.
If implemented properly~\cite{DBLP:conf/uss/MoghimiBHPS20}, constant-time code also helps against single-stepping attacks, as the instructions that are executed and the observable memory accesses are always the same and independent of the secret.

A related approach is randomization, where independent noise is added to decorrelate the observed trace from the secret inputs. However, this only works for certain algorithms (e.g., exponent blinding for RSA) and requires high manual effort. There is also a risk of leaking the random value while applying it to the input or removing it from the output.

\subsubsection{Oblivious code and data patterns}
While constant-time code reliably hides secret data from the attacker, it is trivial to extract the control flow and use that to infer which algorithms are currently running.
In order to also hide the executed code, \emph{oblivious} control flow and data access patterns are necessary.
We define oblivious control flow as follows:
Given a program that is divided into a set of single-entry/single-exit \emph{blocks}, the attacker cannot distinguish which block is currently executed.
This directly requires that an attacker also cannot learn the block's identity via monitoring the data accessed by it.
In contrast to constant control flow, code rewritten to have oblivious control flow may still contain secret-dependent branches, but those are invisible to the attacker.

The efficacy of oblivious code against attacks depends on the block structure and the surrounding controller logic.
For example, due to their low temporal resolution, typical cache attacks cannot distinguish blocks whose size does not exceed the size of a cache line, even if the blocks have slightly varying instruction counts.
However, with more powerful single-stepping attacks, the attacker can precisely count the instructions in a block and even measure their individual latency.
To combat such attacks, the blocks must be carefully crafted to exhibit a uniform profile (\Cref{sec:design:approach:uniform-code-blocks}).

\subsubsection{Ciphertext freshness}
\label{sec:design:sw-leakage-defenses:ciphertext-freshness}
Ciphertext side-channel leakage is caused by deterministic memory encryption in TEEs.
When secrets are repeatedly written to the same address (e.g., in a typical constant-time swap pattern), the attacker can learn those by observing whether the ciphertext changes.
By forcing freshness of the ciphertexts on each write, the leakage can be eliminated.
There are three ways of introducing freshness on software level: First, one could XOR the data before each write with a random mask, such that the ciphertext is independent of the secret. Second, one could interleave the data with random fresh values (e.g., counters) which are updated on every write. A third method is address rotation, which takes advantage of the address-dependent tweak values used in memory encryption. When a variable is copied to a new memory location on each write, it results in a new ciphertext even if the variable's contents are not changed. All methods require heavy instrumentation and bookkeeping to automatically apply them to existing programs~\cite{DBLP:conf/uss/WichelmannPW023}.
Note that ciphertext freshness is only necessary for secret data, as the code usually is written to a random memory location once and does not change during runtime.

\subsection{Obfuscuro}
\label{sec:defenses:obfuscuro-attack}
A solution that partially addresses the issue of code inference is Obfuscuro~\cite{DBLP:conf/ndss/AhmadJXZSL19}.
However, it is susceptible to single-stepping and ciphertext side-channel attacks.
Following the oblivious code approach, Obfuscuro greedily divides the program into a number of code blocks which start with a memory access and have a fixed length of 64 bytes, which is the size of an x86 cache line. The code blocks are fetched from a Path ORAM and copied into a fixed memory region called \emph{scratchpad}. Data obliviousness is achieved through another Path ORAM, which holds 64-byte data blocks. For each code block, exactly one data block is fetched from the ORAM and copied into a data scratchpad. This way, a cache side-channel attacker only sees accesses to oblivious memory and the fixed scratchpads.
Though the code blocks generated by Obfuscuro may have varying instruction counts and latencies, the limited temporal resolution of a cache attack prevents the attacker from effectively exploiting this.

\begin{figure*}[ht!]
    \centering
    \includegraphics[width=0.85\textwidth]{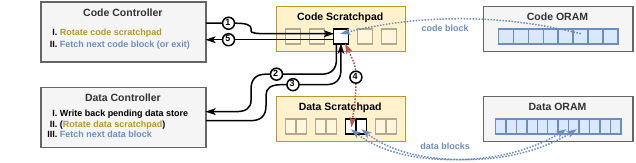}
    \caption{\tool{} execution engine overview. The code controller fetches the next code block and starts executing it (\ding{192}). Then, possibly multiple times, the code block may request a certain memory address from the data controller (\ding{193} \ding{194}) and access it~(\ding{195}). Finally, the code block redirects the execution back to the code controller to fetch the next block (\ding{196}).}
    \label{fig:obelix}
\end{figure*}

\subsubsection{Single-stepping attack}
With single-stepping, temporal resolution is vastly increased, and the attacker can distinguish blocks by counting instructions.
The instruction counts allow to assign labels to individual code blocks, helping identification of branches and the underlying algorithm. If the algorithm is partially known and contains secret-dependent branches, the attacker may be able to extract secret information. Further information is gained by additionally measuring instruction latencies. We evaluate the observable latencies in \Cref{sec:evaluation:security} and show that even for fixed instruction counts, leakage persists.

\subsubsection{Other issues}
Besides the vulnerability to single-stepping, Obfuscuro has a number of other issues. First, it only focuses on Intel SGX, which was complemented by VM-based TEEs after the paper's publication. Thus newer threat models are missing, e.g., it does not take into account ciphertext side-channels, which can leak the identity of a code block even quicker (\Cref{sec:design:approach:ciphertext-protection}). Then, it makes several assumptions that greatly simplify implementation complexity but reduce practical relevance of the security and performance evaluation. For example, Obfuscuro makes a single memory access at the beginning of a block, which the attacker can distinguish as being a load or store by monitoring the instruction latency, or, if possible, by checking whether the ciphertext of the data scratchpad changes. Another restriction is lack of support for position-independent code, which is non-trivial due to code being copied to a scratchpad before execution, breaking relative branches.

To summarize, while the Obfuscuro approach is a good first step at side-channel proof code obfuscation in TEEs, it does not protect against strong adversaries and has several practical limitations.

\section{\tool{} Design}
\label{sec:design}

In the taxonomy of countermeasures in \Cref{sec:defenses} we observed that there is a lack of hardware-based mitigations, and typical software techniques only protect against specific attack classes. We show how we can combine common mitigation approaches in a single drop-in tool, called \tool{}, which defends against all attack classes through specifically hardened oblivious code. In this section, we describe the general design of \tool{}, which includes its execution model, realization of oblivious accesses and defenses against the different attacks. Then, in \Cref{sec:uniform-code-blocks}, we show how we can generate uniform code blocks which are indistinguishable by the attacker. Finally, in \Cref{sec:implementation}, we %
explain how we can apply \tool{} to a program.

\subsection{Execution Model}
The core of \tool{} is its execution engine, which is depicted in \Cref{fig:obelix}. %
Its design is inspired by Obfuscuro~\cite{DBLP:conf/ndss/AhmadJXZSL19}, but has several key conceptual and technical differences. %
During compile time, the program is decomposed into structurally uniform code blocks, which are intended to be indistinguishable by a side-channel attacker. Similarly, data is divided into equally sized blocks, which can happen transparently at runtime.

We use two ORAMs for storing code and data blocks, respectively. Each ORAM is managed by a dedicated controller. When execution enters the code controller, it fetches the next code block from the code ORAM, copies it into the code scratchpad, and \ding{192} starts executing it.
If the code block wants to read or write data, it \ding{193} jumps into the data controller, which copies the desired memory location into the data scratchpad and \ding{194} jumps back.
Then, the code block \ding{195} accesses the data in the data scratchpad.
If the code block does not need data at the given time, dummy data is copied to the scratchpad instead.
Each request to the data controller has a flag marking whether the code block wants to write to the given memory location; if that flag is set for the \emph{previously} fetched data, the data scratchpad will be written back into the data ORAM first.
Finally, execution \ding{196} leaves the code block and jumps to the code controller, in order to fetch the next block.
Given a suitable ORAM implementation and constant-time controllers, a timing attacker gains no direct information about the original location of the accessed code or data.
The code and data scratchpads themselves are protected by the TEE's memory encryption. This way, we achieve security against timing attackers trying to exfiltrate data or infer the executed code.

\paragraph{Multiple data accesses per block}
\label{sec:design:approach:multiple-accesses-per-block}
Steps \ding{193} to \ding{195} may be repeated several times within a code block.
If the instrumented application wants to store data, multiple data controller accesses per code block become mandatory, as we cannot interchange loads and stores at the same offset within a code block without leaking the nature of the access to a timing attacker.
Conveniently, this design can also better represent the load-to-store ratio, which typically heavily favors loads, reducing the number of total executed code blocks and thus expensive code ORAM queries.
To be indistinguishable by an attacker, code blocks are generated in a way that they always have the same sequence of loads and stores.

\subsection{Choosing an ORAM Engine}
\label{sec:design:approach:oram-engine}
Most ORAM schemes described in literature are designed for handling millions of entries in large databases stored on an external server. In addition, they often assume that the ORAM controller (client) is trusted, i.e., that the attacker only resides on the server/network side. 
For TEEs, due to their vulnerability to various local side-channel attacks, the ORAM controller is observable by the adversary as well.
The controller thus needs to be suitably hardened, likely by implementing it in a constant-time fashion. In the following, we briefly discuss two popular ORAM schemes, Linear ORAM and Path ORAM, and which approach we take for our code and data controllers.

A simple and secure approach is \emph{Linear ORAM}~\cite{DBLP:journals/jacm/GoldreichO96}, %
where we iterate over the entire data set and retrieve the desired data via a constant-time selection primitive. Let $N$ be the number of blocks. As we access all blocks, the complexity of Linear ORAM is $\mathcal{O}(N)$. Note that there are numerous possible optimizations, like dividing highly repetitive data (e.g., uniform code blocks) into arrays of indexes into pre-computed maps, achieving compression without compromising on security.

Another simple but asymptotically much faster approach is \emph{Path ORAM}~\cite{DBLP:journals/jacm/StefanovDSCFRYD18}, with a complexity of only around $\mathcal{O}(\log N)$. However, this protocol and others building on top of it only apply to the aforementioned client/server scenario, and the client is inherently non-constant-time. Most problems are caused by the path fetch, where an entire path is copied into a local buffer, and then written back into the tree. For ORAM bucket size $B$, the fetch results in $B\log N$ block accesses and writes. When writing back the path, we must not leak which entries are written back, which are copied into the stash, and which nodes are filled with dummy data. This means that we need to touch every buffer entry for every path node in the worst case, which results in $\left(B\log N\right)^2$ accesses. For a recommended bucket size of $B=4$~\cite{DBLP:journals/jacm/StefanovDSCFRYD18}, this means that the break-even point to Linear ORAM is at around $N\approx 1900$, not counting other factors like bad spatial locality of accesses, the high cost of frequent memory stores, and the linear scan of the position map.

\paragraph{Data controller}
For each data access, we need to always fetch \emph{two} adjacent blocks, the first being the one actually pointed at and the second for supporting an unaligned access to the first. The block size $S_\mathrm{data}$ is bounded by $8\leq S_\mathrm{data}\leq 32$, as the maximum non-vectorized memory access width on x86 is 8 bytes, and two blocks must fit into a single 64-byte cache line to not leak unaligned accesses to a cache attacker. For these parameters and reasonable $N$, Linear ORAM outperforms Path ORAM by a wide margin.

\paragraph{Code controller}
Code blocks have a much larger size $S_\mathrm{code}\geq 64$, making Path ORAM more attractive. However, as explained above, $N$ needs to be sufficiently large to overcome the penalty from the frequent path write-backs and memory stores. In our evaluation, we did not encounter a program where Path ORAM performed well, so we use an optimized Linear ORAM for the code controller as well. %

\subsection{Protecting Against Single-Stepping}
\label{sec:design:approach:uniform-code-blocks}
Unlike typical timing attacks (e.g., cache side-channels), single-stepping offers a very fine-grained view of execution. First, an attacker can simply single-step each code block and count instructions~\cite{DBLP:conf/uss/MoghimiBHPS20}. As we showed in \Cref{sec:defenses:obfuscuro-attack}, filling each block until a fixed size (64 bytes) is reached leads to easily distinguishable code blocks.
In \tool{}, we address instruction counting by dividing each code block into a fixed number of \emph{instruction slots}. Moreover, we ensure that loads and stores are always placed at the same offsets within a block's instruction list. This way, the attacker always observes the same instruction counts.

A stronger attack is additionally measuring the latency of every executed instruction~\cite{DBLP:conf/ccs/BulckPS18}. The resulting latency profile of a code block leads to a unique fingerprint, allowing the attacker to reconstruct a sequence of executed blocks. This sequence reveals the structure of the control flow, which may in turn expose the underlying algorithm or allow inferring secret data if the original implementation was not constant-time. \tool{} assigns each instruction slot a certain \emph{instruction class}, which only contains instructions that are indistinguishable for the given measurement setup.

However, it is possible that a fixed latency pattern is still insufficient for full protection against single-stepping attackers: In some cases, the alignment of an instruction influences its measured latency due to peculiarities of the CPU frontend~\cite{DBLP:conf/uss/PudduSHC21}. 
This can be avoided by placing each instruction slot at a fixed alignment, and filling the remainder of each slot with an instruction that does not expose the aforementioned latency variations, like a multi-byte no-op. %

\subsection{Preventing Ciphertext Side-Channel Attacks}
\label{sec:design:approach:ciphertext-protection}
With the previously described countermeasures, even precise timing measurements cannot distinguish individual code blocks. However, several current and proposed full-VM TEEs employ deterministic memory encryption, i.e., at a fixed memory location a given plaintext always yields the same ciphertext. Deterministic encryption is problematic for the code and data scratchpads, as an attacker can easily assign a label to each executed code block by observing the ciphertext of the code scratchpad. Similarly, the attacker can track data blocks by keeping label/ciphertext pairs, and updating the ciphertext part whenever they notice a change of the data scratchpad following a store from the code block.

As described in \Cref{sec:design:sw-leakage-defenses:ciphertext-freshness}, there are three methods for ensuring ciphertext freshness: Adding a random mask, interleaving with a counter, and rotating store addresses.

\subsubsection{Protecting code}
For the code scratchpad, masking is not an option, as code must be provided in plain form to be executable by the processor.
Interleaving is disadvantageous as well, as the code would need to be modified to jump over or incorporate the counters, reducing efficiency of code blocks. 
The last option, rotating the location of the code scratchpad for each block, has neither of those disadvantages and is thus best suited for protecting the executed code.
While there may be eventual repetition of code ciphertexts, a sufficiently large pool of locations greatly reduces the probability that a certain code block ends up twice at the same address~\cite{DBLP:conf/sp/LiWW0TZ22}. Aside from that, we found that picking a different code scratchpad location for each next block actually improves performance due to decreasing machine clears from self-modifying code (see \Cref{sec:discussion:smc}).

\subsubsection{Protecting data}
Since we allow the program to modify data, we need to protect both the scratchpad and the ORAM.
Masking and interleaving both increase memory usage by 100\%
However, rotating addresses of large memory objects is even more costly, so it is only suitable for small regions, i.e., the data scratchpad.
Effective masking requires a continuous supply of randomness, which was found to be quite costly~\cite{DBLP:conf/uss/WichelmannPW023}.
For this reason, we recommend protecting larger amounts of data with interleaving. 
In a Linear ORAM implementation, this means splitting up data into chunks of $\frac{E}{2}$ bytes for encryption block size $E$, and putting counter values of $\frac{E}{2}$ bytes in between, such that each encryption block consists of a data chunk and a counter. %
Each time we write back into the ORAM, we increment every counter, so we get new ciphertexts for every ORAM entry.
Interleaving guarantees that a ciphertext can only repeat after at least $2^{8\cdot\frac{E}{2}}=2^{4E}$ iterations (i.e., $2^{64}$ for 16-byte encryption blocks), contrary to masking, where a fast non-cryptographic random number generator may lead to accidental collisions.

\section{Single-Stepping Resistant Code Blocks}
\label{sec:uniform-code-blocks}

As shown in the execution engine overview in \Cref{fig:obelix}, we divide the code into same-sized blocks, which are obliviously fetched and copied into a fixed code scratchpad at runtime. While this protects against timing attackers with limited resolution, single-stepping offers instruction granularity and hence allows precisely counting and measuring the instructions executed within a block.
We thus need to devise a way to generate uniform blocks, which exhibit a fixed structure and latency pattern that even a single-stepping attacker cannot distinguish. At the same time, the uniform blocks should not increase performance overhead compared to the na\"ive greedy block creation approach.

There are two major ways to approach this. First, one could modify the instruction scheduler in the compiler to emit a suitable instruction sequence, in a similar approach as for Very Large Instruction Word (VLIW) architectures. However, this requires writing a new instruction scheduler or at least heavily modifying the existing one, making this approach very complex and hard to maintain. We thus picked the other approach, which instead of ensuring that the generated machine code conforms to a certain pattern, computes a \emph{block pattern} that is optimized for the existing machine code. The machine code is then post-processed and divided into blocks in an instrumentation pass.

To achieve this, we first show how we can group instructions into classes, which contain instructions mutually indistinguishable from each other by single-stepping. We then first discuss the high-level layout of code blocks, and conclude with the uniform pattern generation algorithm.

\subsection{Classifying Instructions}
\label{sec:uniform-code-blocks:instruction-analysis}
As a first step, we need to determine which instructions fall into the same classes, i.e., can be used interchangeably without leaking their identity to the single-stepping attacker. To minimize the number of classes and avoid a lot of dummy instructions per block, we focus on the base instruction set, i.e., scalar arithmetic and memory accesses, and disable more volatile extensions like vector arithmetic. The general purpose instructions cover all common use cases.

\subsubsection{Measurement setup}
\begin{figure}[t]
    \centering
    \includegraphics[width=0.25\textwidth]{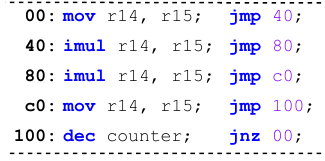}
    \caption{Microbenchmark for the \texttt{mov} and \texttt{imul} instructions. We execute each instruction two times in an ABBA pattern, always aligned to a cache line, to avoid bias from other CPU components.}
    \label{fig:microbenchmark}
\end{figure}

While resources like \url{uops.info} offer very precise latency information, this kind of precision is in practice not achievable by an attacker, who has a lower temporal resolution due to noise from the enclave entry/exit context switches. We thus devise an own microbenchmark specifically for single-stepping.
For each pair of instructions, we generate a standardized gadget which is designed to minimize noise from other CPU components, and measure it \num{1000000} times.
An example gadget for the x86 \texttt{mov} and \texttt{imul} instructions is shown in \Cref{fig:microbenchmark}.

\subsubsection{Results}

\begin{figure*}[ht!]
    \centering
    \begin{subfigure}[]{0.3\textwidth}
        \centering
        \includegraphics[width=\textwidth]{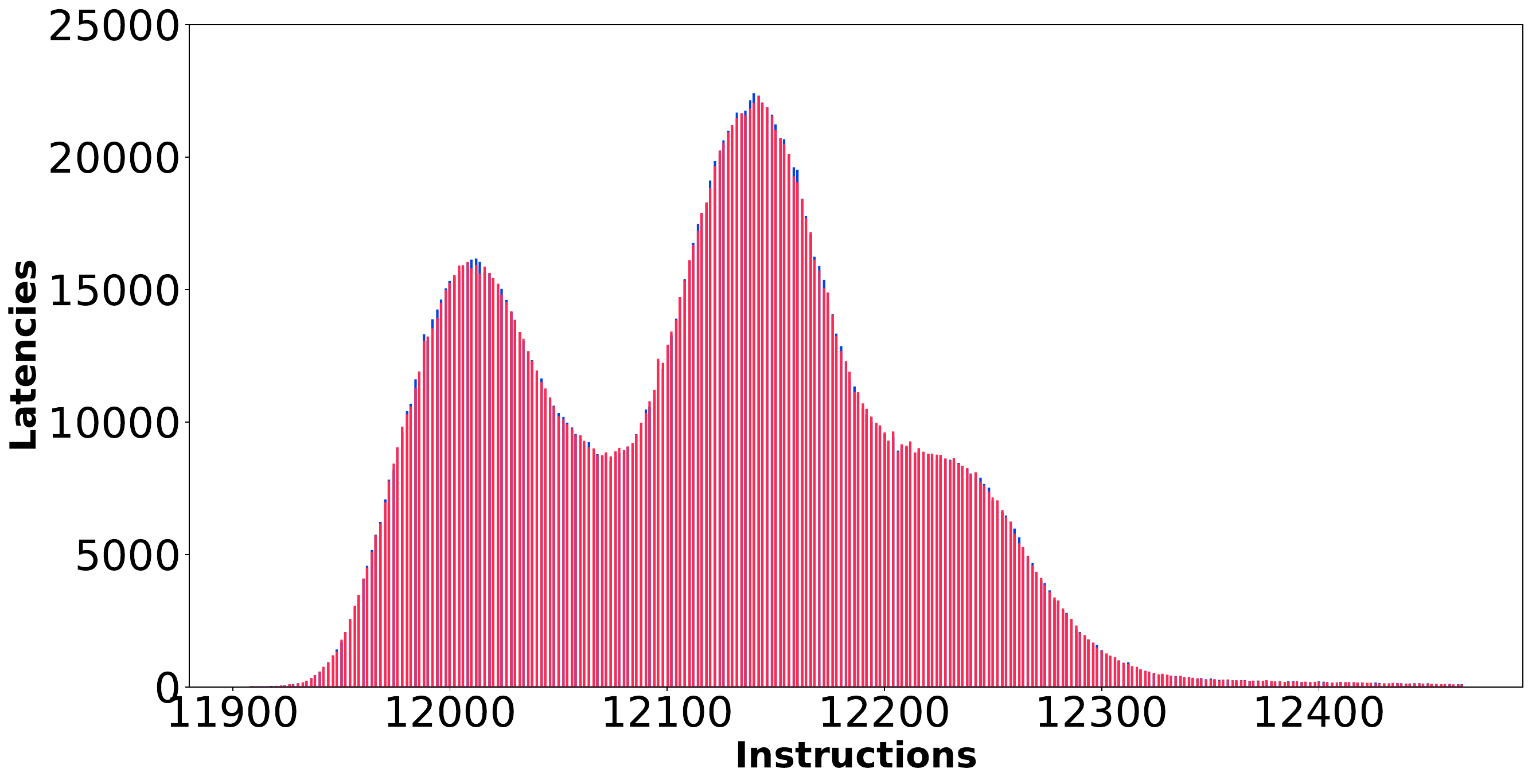}
        \caption{Histogram of \texttt{mov} vs. \texttt{imul}}
        \label{fig:instruction-analysis:imul-hist}
    \end{subfigure}
    \hfill
    \begin{subfigure}[]{0.3\textwidth}
        \centering
        \includegraphics[width=\textwidth]{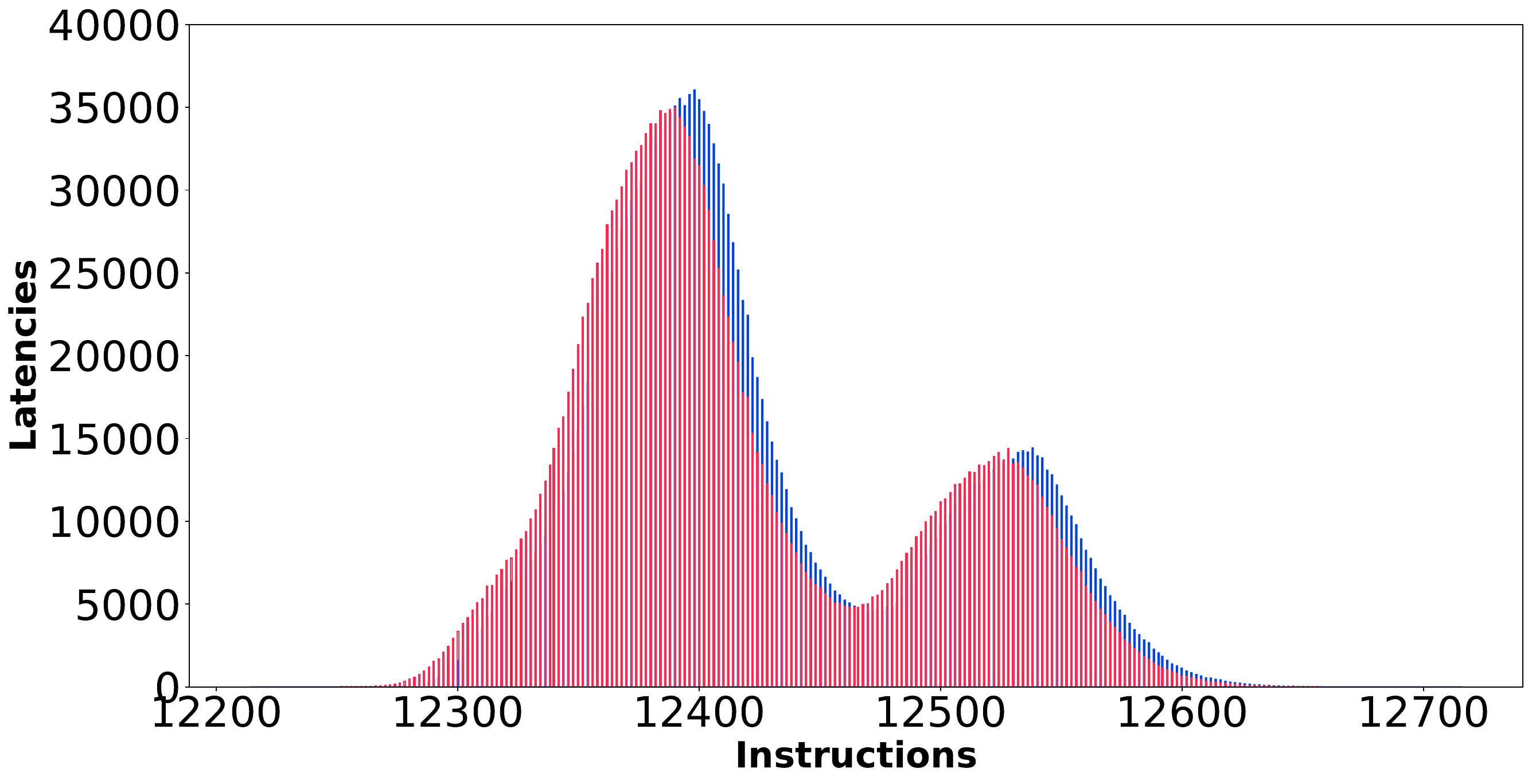}
        \caption{Histogram of \texttt{mov} vs. \texttt{div}}
        \label{fig:instruction-analysis:div-hist}
    \end{subfigure}
    \hfill
    \begin{subfigure}[]{0.3\textwidth}
        \centering
        \includegraphics[width=\textwidth]{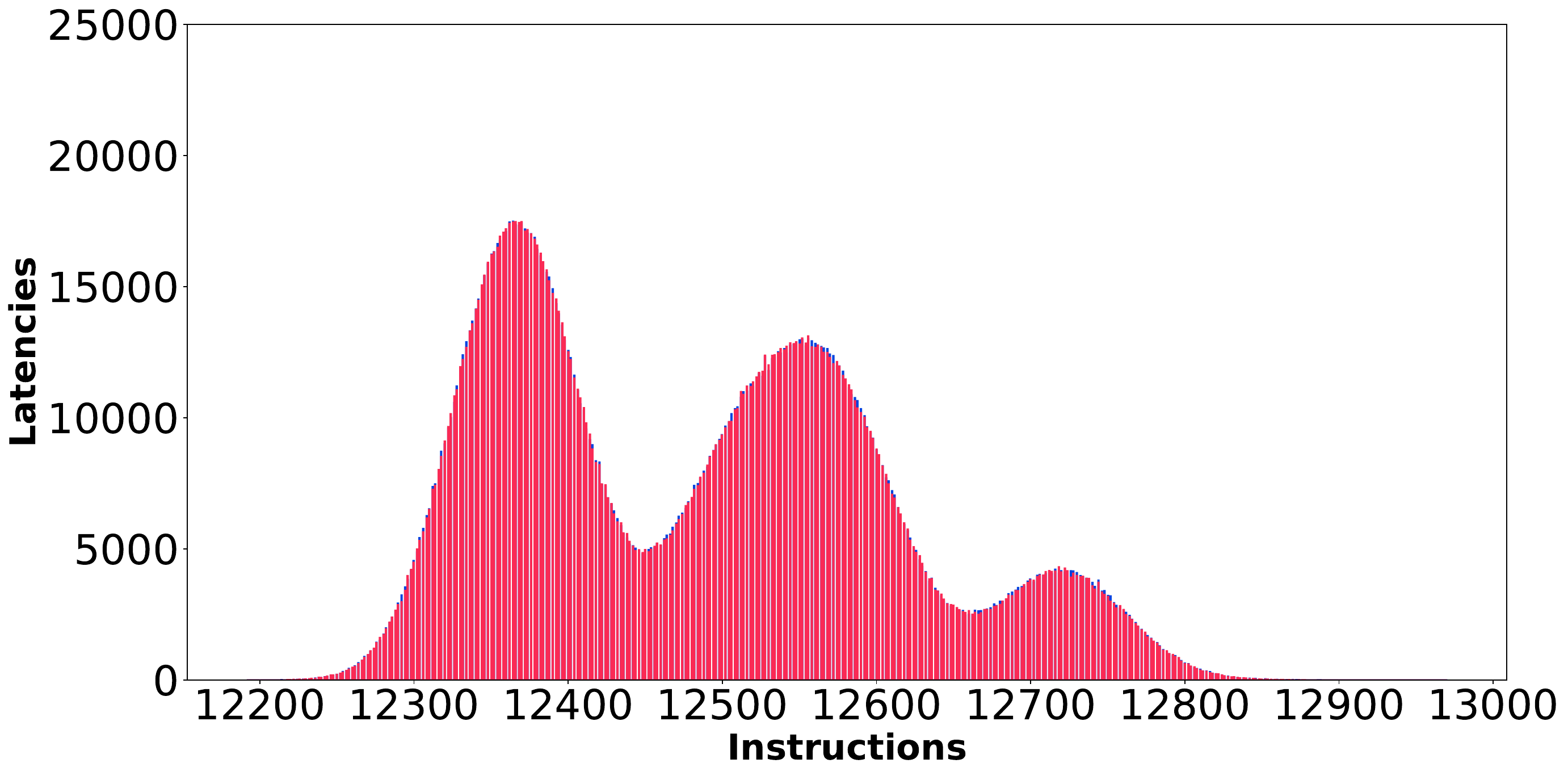}
        \caption{Histogram of \texttt{inc} vs. \texttt{inc}}
        \label{fig:instruction-analysis:inc-hist}
    \end{subfigure}

    \begin{subfigure}[]{0.3\textwidth}
        \centering
        \includegraphics[height=0.15\textheight]{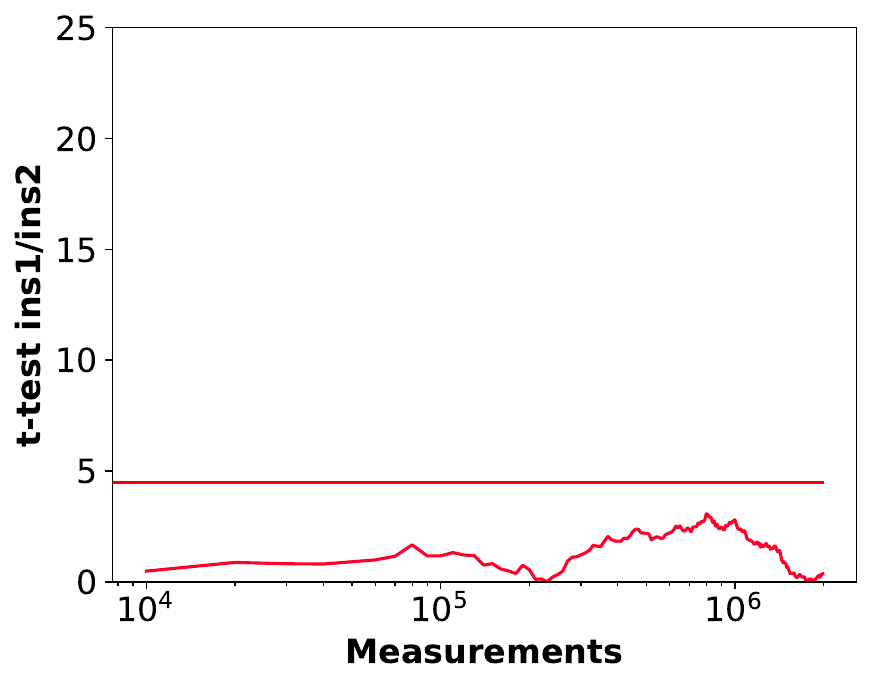}
        \caption{$t$-test progression of \texttt{mov} vs. \texttt{imul}}
        \label{fig:instruction-analysis:imul-t}
    \end{subfigure}
    \hfill
    \begin{subfigure}[]{0.3\textwidth}
        \centering
        \includegraphics[height=0.15\textheight]{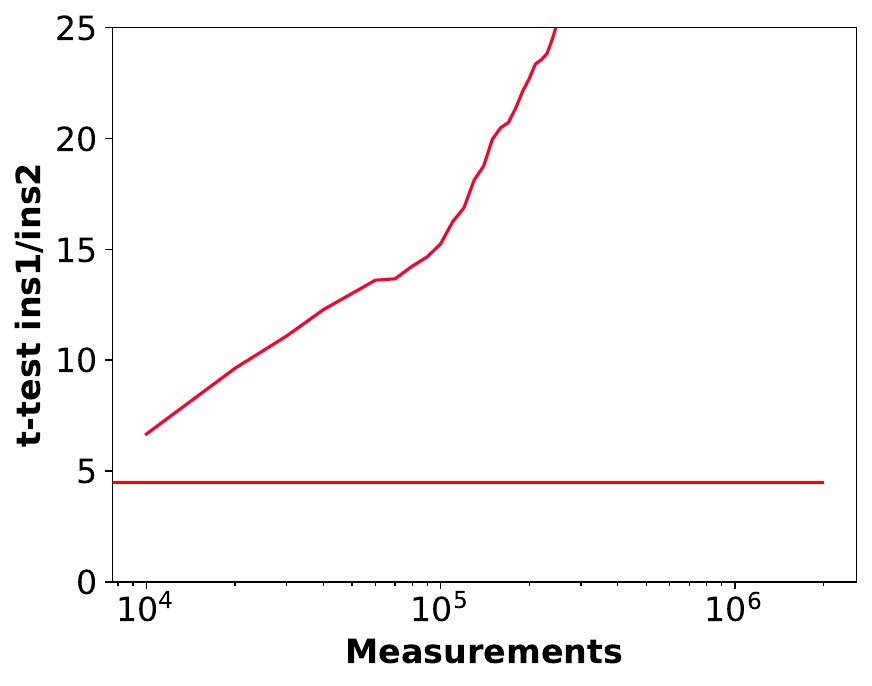}
        \caption{$t$-test progression of \texttt{mov} vs. \texttt{div}}
        \label{fig:instruction-analysis:div-t}
    \end{subfigure}
    \hfill
    \begin{subfigure}[]{0.3\textwidth}
        \centering
        \includegraphics[height=0.15\textheight]{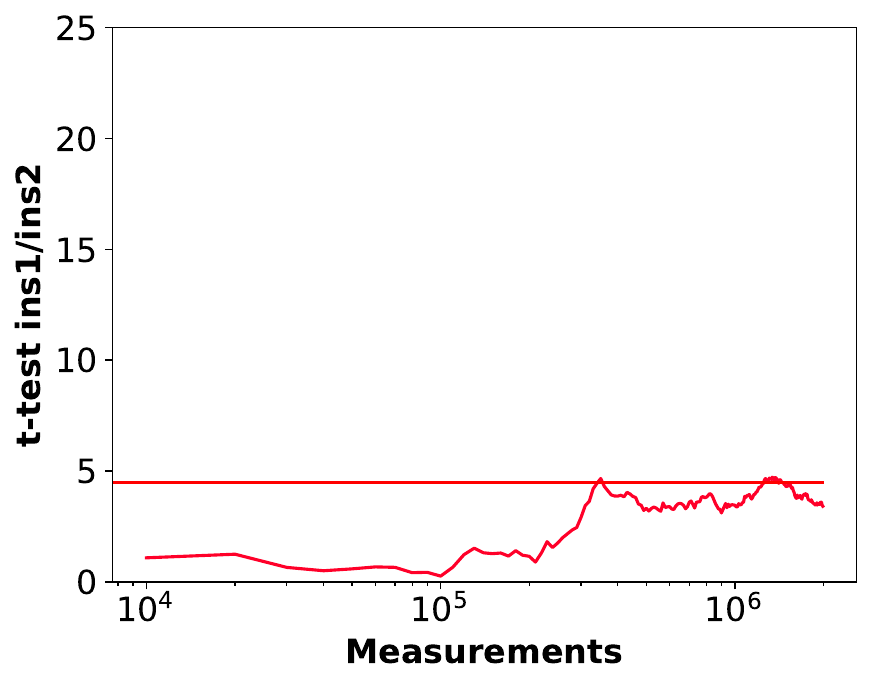}
        \caption{$t$-test progression of \texttt{inc} vs. \texttt{inc}}
        \label{fig:instruction-analysis:inc-t}
    \end{subfigure}
    
    \caption{Histograms and $t$-test progressions for some instruction latency experiments on Intel SGX.
    }
    \label{fig:instruction-analysis}
\end{figure*}

After composing the instruction gadgets, the microbenchmark is loaded into an SGX enclave (respectively, an application running inside an AMD SEV VM). Our test systems are two machines, first an Intel Core i7-9750H CPU (Coffee Lake) with 16 GB of RAM, running Ubuntu
22.04.3 LTS with a custom kernel version 5.9, and an AMD EPYC 7763 (Zen3) with 128 GB of RAM, running Ubuntu 22.04.3 LTS with a custom
kernel 5.19 with SEV-Step patches.
Both systems were thoroughly configured to have as little noise as possible, e.g., by disabling dynamic frequency scaling and memory prefetchers, isolating cores from the kernel scheduler, disabling SMT, and minimizing overall system load.
This way, we emulate a strong attacker who is able to conduct measurements with the maximum possible precision. We single-step the benchmark, recording \num{1000000} samples for every executed instruction.

As visible in \Cref{fig:instruction-analysis}, the Intel SGX instruction measurements are quite noisy and produce multiple peaks in the histograms (for AMD, see \Cref{fig:sev-measurements} in the appendix). Most of the variation is caused by the context switches to/from the enclave, where, among other activities, the entire execution state is saved/restored from memory. As a method to get a quantification of the attacker's capabilities to distinguish the latency distributions of two instructions, we employ Welch's $t$-test with a threshold of $4.5$~\cite{DBLP:journals/jce/SchneiderM16}. If $\abs{t}<4.5$, we conclude that the instructions are not distinguishable for \num{1000000} measurements. To find out whether $t$ converges, we plot its progression with an increasing amount of measurements.

\Cref{fig:instruction-analysis:imul-hist} and \Cref{fig:instruction-analysis:div-hist} show the histograms for a comparison of \texttt{mov~reg,~reg} vs. \texttt{imul~reg,~reg} respectively \texttt{div~reg}. Per \url{uops.info}, on our Coffee Lake CPU, \texttt{mov} has a latency of \num{0.25}, \texttt{imul} a latency of \num{3} to \num{4}, and \texttt{div} a latency of \num{5} to \num{89}.
As is apparent in the histograms, \texttt{imul} is nearly indistinguishable from \texttt{mov}, while \texttt{div} clearly deviates. This is supported by the $t$-tests: For \texttt{imul}, $t$ stays well below the threshold (\Cref{fig:instruction-analysis:imul-t}), while for \texttt{div} it immediately deviates (\Cref{fig:instruction-analysis:div-t}), expressing clear distinguishability by an attacker.
We conducted those measurements for several types of arithmetic instructions, and identified two classes with instructions which were indistinguishable for our sample count: The first one, named \texttt{class1}, contains all standard arithmetic like addition, multiplication and bit shifts, effective address computations and (conditional) register/immediate moves.
The second one, \texttt{class2}, contains only the division instructions.
However, as indicated by the \url{uops.info} measurements, those exhibit an operand-dependent latency range, so they may leak some information on their operands. If this leakage is not tolerable, we recommend avoiding such instructions until security features like Intel DOIT~\cite{intelDoitDocs} become available, which enforce data operand-independent execution times. As a temporary workaround, we could remove and emulate such instructions by common arithmetic.

Note that, while \texttt{mov} and \texttt{imul} turned out indistinguishable in our experiments, they \emph{should} be distinguishable by an attacker doing enough measurements, as they have a slightly different latency. In fact, with far more than the million measurements in our experiment, a difference may eventually become apparent. However, gathering so many measurements for a real target is very difficult in practice, especially if the target deploys application-level mitigations which prevent repeated execution. This is supported by another observation, that noise causes the $t$-test to first signal a leakage, but then drop below the threshold again. This effect is apparent in \Cref{fig:instruction-analysis:imul-t}, where $t$ briefly approaches $4.5$ and then drops to almost $0$ again. In our sanity check where we compared two identical \texttt{inc} instructions (\Cref{fig:instruction-analysis:inc-hist} and \Cref{fig:instruction-analysis:inc-t}), the $t$-test falsely signaled a leakage.

We conclude that, while it is impossible to \emph{prove} that an attacker cannot distinguish two instructions from the identified equivalence classes if given an even better optimized experiment environment and sufficient measurements, we deem it very unlikely in practice. 
If the user decides that this risk is too high, they can still increase the number of classes and narrow the latency interval of the contained instructions.

\begin{figure*}[ht!]
    \centering
    \includegraphics[width=0.85\textwidth]{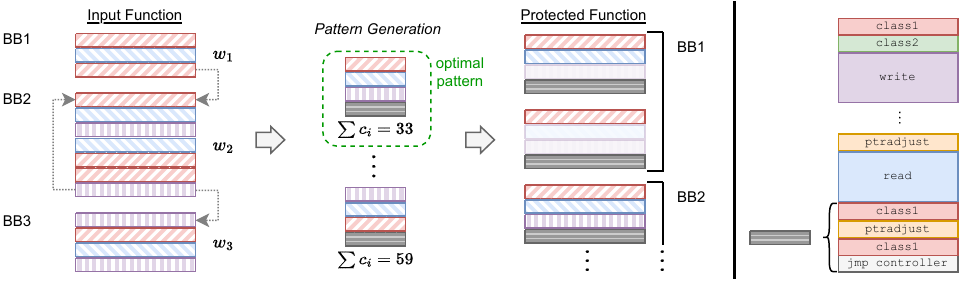}
    \caption{The function instrumentation process (left) and an example code block (right). Using the basic block profile from the input function, we compute costs for different pattern candidates, which are generated by a genetic algorithm. The basic blocks are then translated according to the optimal pattern, yielding a protected version of the function.}
    \label{fig:pattern-generation}
\end{figure*}

\subsection{High-Level Block Layout}
A code block always executes from its beginning, and concludes with an unconditional jump to the code controller at its end. The exit jump is preceded by a single (conditional) move instruction which determines the address of the next block to execute. Between the entry and exit points the block can have an arbitrary sequence of application instructions and memory accesses.

\subsubsection{Instruction slots}
To ensure a fixed block length and avoid leakage due to varying instruction alignment, we partition each code block into a number of \emph{instruction slots}, as explained in \Cref{sec:design:approach:uniform-code-blocks}. In practice, every instruction slot has a size of 8 bytes and holds one instruction of up to 7 bytes and one multi-byte no-op filling up the remainder of the slot. Most instructions from the x86 base instruction set fit into 7 bytes; notable exceptions are moves with a 64-bit immediate and address expressions with 32-bit displacements, which we split into a sequence of equivalent shorter instructions before instrumentation. We verified that the multi-byte no-op instructions recommended by the CPU vendors are indistinguishable through single-stepping.

\subsubsection{Position-independent code}
Modern programs usually employ position-independent code, which means that the code and data can be mapped to an arbitrary virtual address during startup.
To access a global variable on x86-64, a \emph{RIP-relative addressing mode} is used, i.e., the address is computed dependent on the current instruction pointer. However, this becomes a problem when the code is dynamically copied into a scratchpad at runtime. We handle this by introducing a special instruction class, called \texttt{ptradjust}, which adds the difference between the addresses of the code scratchpad and the original code to the pointer.
If an instruction uses a RIP-relative addressing mode, it is translated to be followed by such a pointer adjustment.

\subsubsection{Memory accesses}
A code block may contain multiple memory accesses. Consistent with the uniformity requirement, each access is located at a fixed position within the block and always performs the same type of memory access (load or store). A memory access consists of the following components: We first need to load the address of the memory location we want to access, and the offset to which we want to return in the code block. Both are stored in general-purpose registers reserved for the code and data controllers, as we want to avoid polluting the application's stack. We then jump into the data controller, which performs the ORAM fetch and copies the desired memory into the data scratchpad. Subsequently, we execute the memory access instruction, where we have replaced the address operand by an access to our data scratchpad.
With careful optimization, we managed to fit a memory access into 24 bytes, i.e., three instruction slots. We also ensured that all instructions involved in a memory access have a consistent latency.

\subsubsection{Code block size}
The code block exit point is directly preceded by three fixed instruction slots, which are needed for emulating conditional jumps. Thus, with the exit jump, the code block end takes four instruction slots. If we assume that a function both loads and stores data, we need another 6 slots for memory accesses. This means that without any other instructions any block already needs 10 slots, which are 80 bytes.
The code block size directly impacts performance, as larger blocks mean fewer costly code ORAM fetches, but may also come with an increased number of dummy memory accesses. Security is also affected: In theory, the block size and pattern can be selected in a way that the function's biggest basic block can be encoded into a single code block. However, this may leak a lot about the function's internal structure, so smaller blocks are desirable.
We found that a block size of 160 bytes (20 instruction slots) is a reasonable upper bound. %

\subsection{Generating an Efficient Block Pattern}
\label{sec:uniform-code-blocks:pattern-generation}
Given the overall block structure, we now need to determine a good sequence of \texttt{classX} instructions and memory accesses. For this, we first compute profiles for the protected functions. These are then used in the second step to estimate the runtime cost of converting a function to each of the generated code block pattern candidates, so we can choose an optimal candidate in the end. The instrumentation process is illustrated in \Cref{fig:pattern-generation}.

\subsubsection{Getting profiles for basic blocks}
To generate a profile for a function, we traverse its basic blocks and then their machine instructions,
assigning each one of the following classes: \texttt{load} (memory reads), \texttt{store} (memory writes), \texttt{ptradjust} (for converting RIP-relative pointers to absolute ones), and \texttt{class*} for the various instruction latency classes.
Basic blocks are also given a weight, which is higher if they appear in a loop or child function.

\subsubsection{Cost function}
To estimate the efficiency of any pattern candidate, we devised a function that takes a pattern candidate and the weighted basic block profiles, and then simulates the basic block translation as it would occur in the actual instrumentation. Nearly all runtime cost is caused by ORAM fetches, so we want to both minimize the number of executed code blocks and dummy memory accesses. For each basic block $i$, we thus count the number of resulting code blocks $b_i$ and the number of dummy loads $l_i$ and stores $s_i$ which need to be generated due to mismatches between the original code and the block pattern. Given weight $w_i$, the cost for instrumenting basic block $i$ is then computed as $\mathrm{cost}(i)=w_i\cdot(b_i+l_i+2 s_i)$. Stores are weighted double due to the higher runtime cost of writing to memory. That cost function balances the number of code blocks and memory accesses, but may be adjusted to better reflect workloads that have lots of code or handle large amounts of data.

\subsubsection{Computing the pattern}
Given the cost function, we can now look for a pattern that minimizes the estimated runtime cost when applied to the given basic blocks. %
The pattern search has a few constraints: First, there is a fixed number of \emph{instruction slots} per code block which need to be filled. The pattern must contain all instruction classes present in the profiles. Then, for good runtime performance, we want to avoid inserting too many dummy instructions. As a final constraint, the algorithm should not take too much time to run, to avoid slowing down the compilation.

\paragraph{Brute-force search}
One approach for finding a suitable pattern is a brute-force search: For example, for a small block size of 96 bytes (12 slots) and both read and write accesses, we need to distribute two memory accesses and two \texttt{classX} instructions. The resulting search space is reasonably small and can be fully scanned easily. However, for larger block sizes, the search space quickly grows, making brute-force search infeasible in the generic setting.

\paragraph{Genetic algorithm}
As a more efficient alternative, we devised a simple genetic algorithm. Initially, we generate a population of $P$ random pattern candidates. Then, all candidates are validated, i.e., we ensure that mandatory instruction classes and the suffix is present. If a mandatory class is missing, we write it into one or more random slots. After computing the cost function for each candidate, we take the top $t$ candidates with the lowest costs and store them in a set $T$. We then create a copy $\Tilde{T}$ which contains slightly mutated versions of the candidates from $T$ with random insertions, deletions and modifications. Subsequently we crossover $T$ and $\Tilde{T}$ to a new population $\Tilde{P}$ by combining random prefixes and suffixes of the respective elements. We then set $P\coloneqq\Tilde{P}\cup T\cup\Tilde{T}$, so we keep both the top candidates in their original and slightly mutated versions. Finally, we replace duplicates by entirely new random candidates and repeat the process.

We limit the genetic search both by number of generations (\num{10000}) and total runtime (10 seconds). The number of top candidates was set to $t=15$, leading to a population size of $\abs{P}=2t^2+2t=480$. We also fine-tuned several other parameters like the mutation rates. In our experiments, we seldom reached the generation cap, usually running a few thousand generations. In fact, for all targets which we evaluated, our algorithm converged within the first few hundred generations, and even running it for a way longer time did not yield a better result.

As a result, we now have a code block pattern that fits the given function and minimizes the number of ORAM queries for dummy accesses.

\section{Implementation}
\label{sec:implementation}

As we work towards a generic TEE protection framework with as little manual intervention as possible, we need to automate almost all instrumentation steps. \tool{} initially asks the user to mark the functions that need protection. Everything else is taken care of by a number of LLVM 17 compiler passes. First, we propagate the function markers to all child functions in the respective call tree. Then, we identify data which is accessed within said call tree, so we can generate initialization calls for the data controller. Finally, after computing a code block pattern for the functions to be instrumented, we rewrite the machine code of the marked functions to fit the pattern, and add a jump to the code controller entry point at the top of the parent function. In total, we added \num{9710} lines to LLVM.

In this section, we describe the aforementioned instrumentation steps and the implementation of the code and data controllers.

\subsection{Marking Functions Needing Protection}

\begin{figure}[t]
    \centering
    \includegraphics[width=0.85\linewidth]{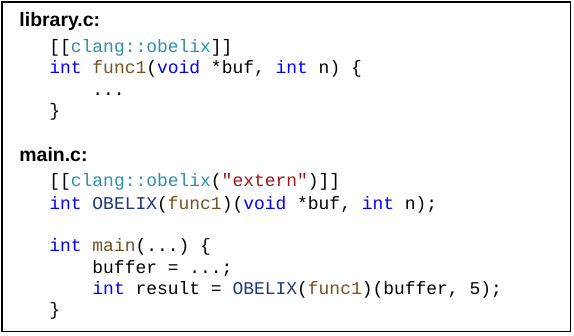}
    \caption{Library function annotated with a custom attribute (C23 syntax) that indicates to the compiler that the function should be protected (top). For calls from the application, the user can annotate the function prototype with a specialized version of the attribute (bottom), allowing the compiler to insert the necessary data ORAM initialization call for the \texttt{buffer} variable. The \texttt{OBELIX} preprocessor macro adjusts the function name to allow the user to use both the original and the protected versions of the \texttt{func1} function.}
    \label{fig:attribute}
\end{figure}

As we only want to protect particularly sensitive portions of the program, we ask the user to mark the corresponding functions with a dedicated attribute (\Cref{fig:attribute}). This is the only manual step required by \tool{}.
Given an annotated parent function, we want to instrument its entire call tree, to avoid leaving secure mode in between and thus leaking control flow.
We cannot use an optimal pattern for every child function, as this would allow the attacker to distinguish them during oblivious execution.
Instead, we need to compute a single pattern for the entire call tree.
To enable this, we create a copy of each child function which is then associated with the call tree parent function. We leave annotations on all copied child functions, such that they are processed together with the parent and later loaded into the same code ORAM.

\paragraph{Calls to external functions}
Calls to functions not residing in the same library are more challenging, and currently not supported by our implementation. The other library may be compiled with a different compiler and/or not support \tool{} at all. Thus, upon encountering a call to an external function, we would need to temporarily switch back from oblivious to normal execution, and resume oblivious execution when the external function returns. This switch would temporarily break obliviousness as the attacker now knows the precise location in the program, and what external function it relies on. By introducing a random number of dummy blocks around the context switch we could again establish a secure state, assuming that there are only very few external calls in the function. Due to the loss of obliviousness and introduction of unprotected code into a secure context we advise against external calls, and recommend moving those outside the protected call tree.

\subsection{Initializing the Data ORAM}
The functions in the protected call tree work with various types of data: They may take pointers to input data, write output buffers, access global values or store local variables on the stack. Those addresses must be present in the data ORAM. We automate this by scanning the function parameters, analyzing the stack frame layout and detecting accesses to global variables, and then generating an ORAM insertion call for each pointer.
This works as long as we can determine the size of each object at compile time, and those objects do not contain pointers to further objects. If we want to support nested objects, we need static points-to and type analysis. At the time of writing, due to a breaking change beginning with LLVM 15 that removed type information from pointers in IR, only older LLVM versions are supported by corresponding static analysis tools.
As a temporary workaround for the missing analysis and to ensure that \tool{} is stable even when a used memory object is not contained in the data ORAM, we created a fallback in the ORAM fetch logic that detects missing addresses and lazily inserts them on demand, at the cost of briefly violating obliviousness due to changing the size of the ORAM.

\subsection{Ciphertext Side-Channel Protection}
To defend against ciphertext side-channel attacks, we use the methods described in \Cref{sec:design:approach:ciphertext-protection}. The entire logic exists in the data and code controllers, and can thus easily be enabled without changes to the code blocks, depending on whether the TEE is vulnerable to ciphertext side-channels or not. We rotate the code scratchpad each time we fetch a block. For our benchmarks, we use a code block size of 160 bytes and ten memory pages (\num{40960} bytes) for hosting the scratchpad. This leads to 160 possible 256-byte aligned locations. The size of the pool of possible scratchpad locations has little impact on performance once it is fully allocated, so it can be easily expanded if desired.

To protect data, we rotate the data scratchpad similar to the code scratchpad, and apply interleaving to the data ORAM. For performance, we chose a data block size of 16 bytes, so the data scratchpad, which holds two blocks, fits in an AVX2 vector register. The data ORAM itself is a single contiguous array of blocks, where the mapping of original pointers to ORAM indices is kept in a separate list. As the encryption block size is 16 bytes, each data block is divided into two halves and interleaved with 8-byte counters, which are incremented on each store. Since we must update both the counter and the data at the same time, we use \texttt{vpunpck*} instructions to merge counter and data into a single vector register, which is then written back. This way, we never store the same plaintext at the same location.

\begin{figure*}[t!]
   \centering
   \begin{subfigure}[]{0.352\textwidth}
       \includegraphics[width=\textwidth]{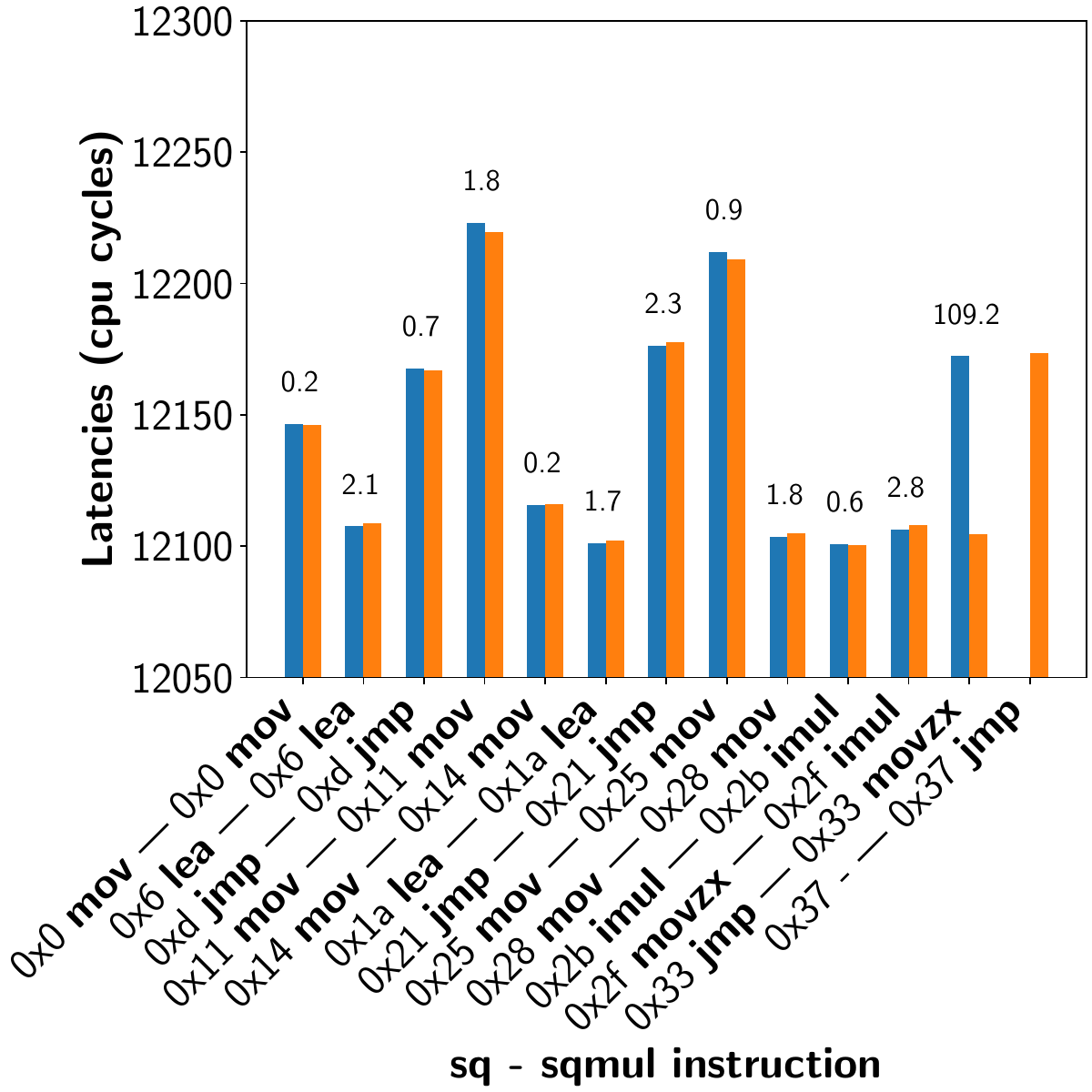}
       \caption{\toolI{}}
       \label{fig:block-comparison:i}
   \end{subfigure}
   \hfill
   \begin{subfigure}[]{0.621\textwidth}
       \includegraphics[width=\textwidth]{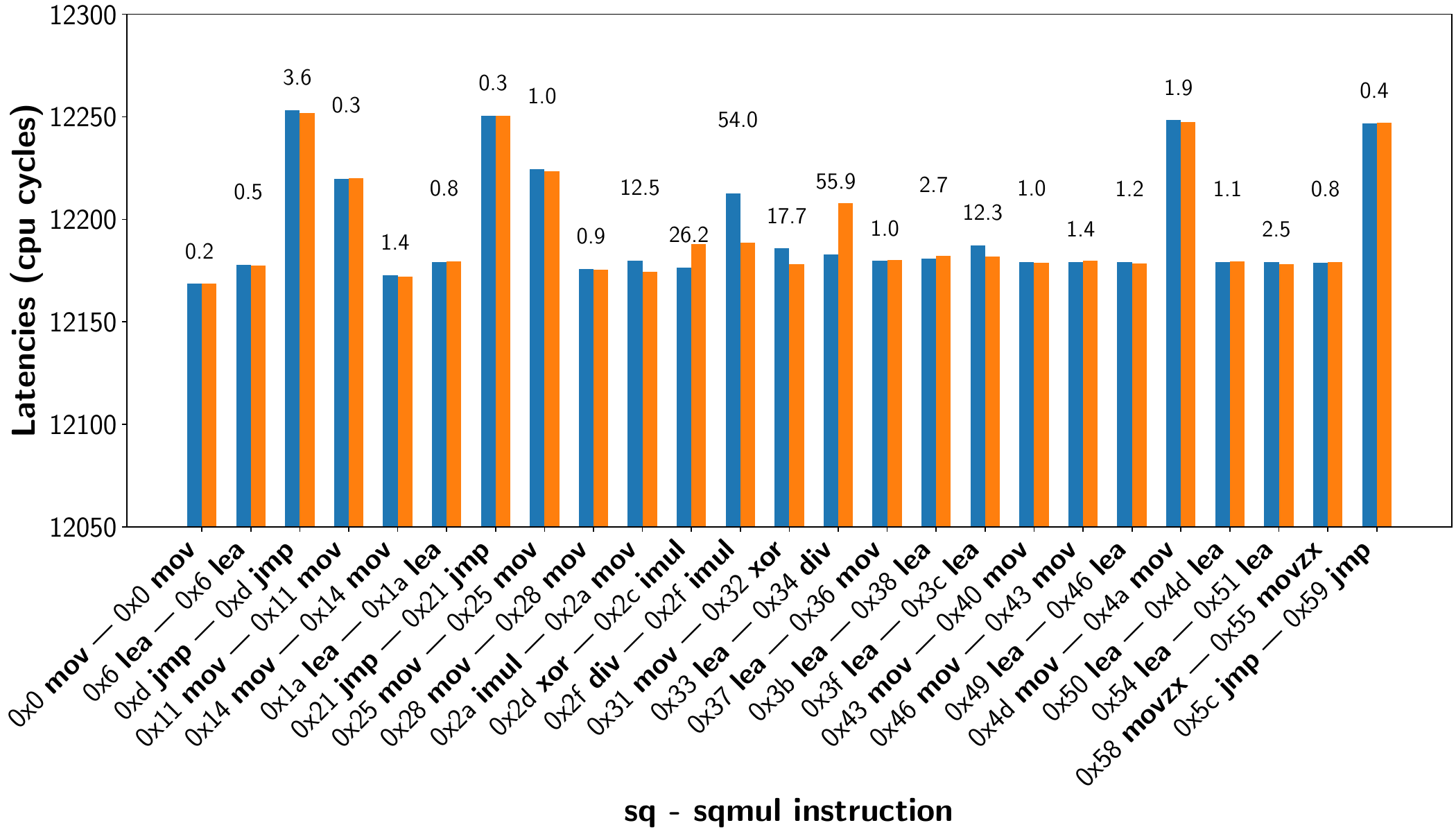}
       \caption{\toolII{}}
       \label{fig:block-comparison:ii}
   \end{subfigure}
   \\
   \begin{subfigure}[]{\textwidth}
       \includegraphics[width=\textwidth]{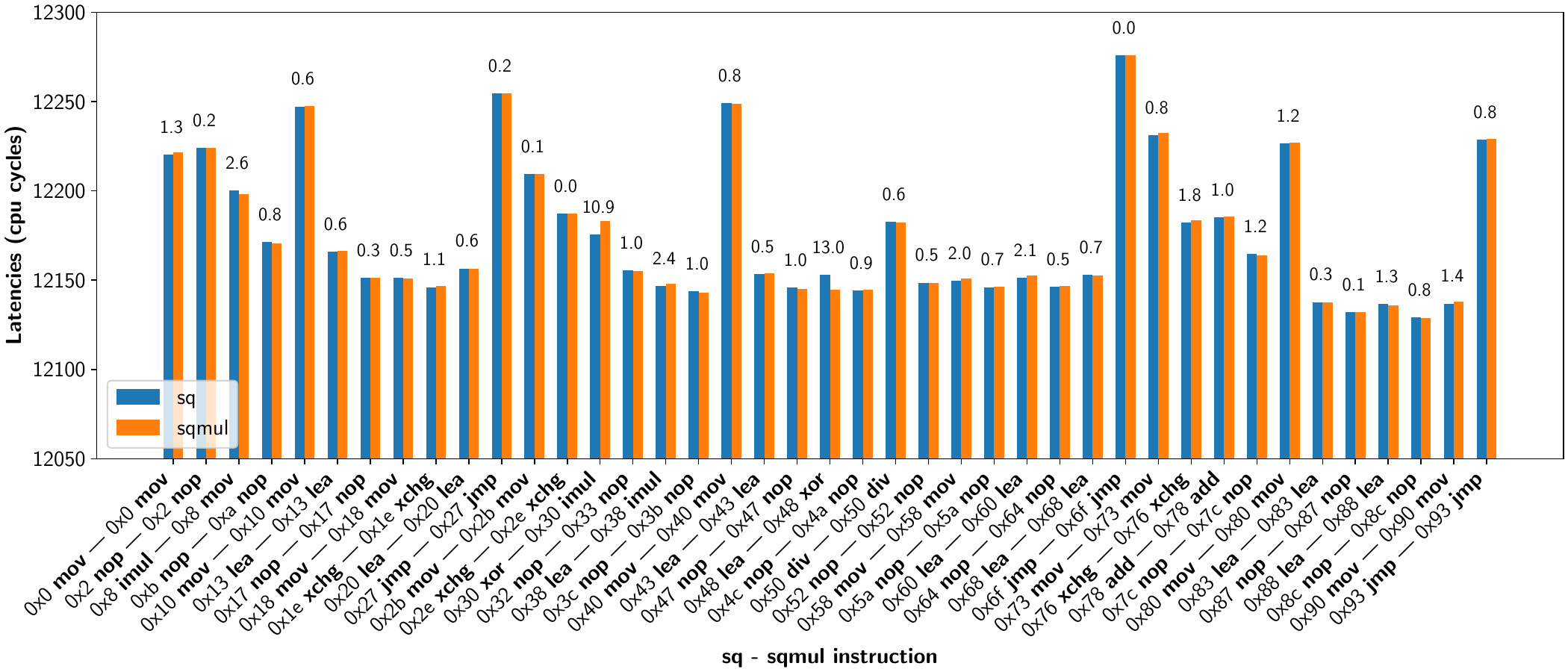}
       \caption{\toolIV{}}
       \label{fig:block-comparison:iv}
   \end{subfigure}
   \caption{Single-stepping measurements of two code blocks from a square \& multiply function. The blue bars represent a ``square, mod'' code block, while the orange bars map to ``square, multiply, mod''. The numbers above the bars are the corresponding $t$-test result.}
   \label{fig:block-comparison}
\end{figure*}

\subsection{Controller Implementation}
We complement the compile-time instrumentation with a runtime library, which contains the code and data controllers and the ORAM implementations. The controller code is a mix of fully constant-time C and assembly code to ensure that it does not leak parts of the protected execution state or the addresses which are fetched from the ORAMs. The ORAM operations are vectorized to maximize throughput, using blending with a mask instead of \texttt{cmov}. We defer oblivious write-back operations until the next fetch, reducing load on the memory system.

\subsection{Correctness}
To ensure that our protection preserves program semantics, every replacement of an instruction must have the same architectural behavior. When entering the controllers, status flags and register values are saved and later restored. %
Finally, dummy instructions are selected to be effective no-ops with zero side-effects: For instruction class \texttt{class1}, the instruction \texttt{add~reg,~0} would have a suitable latency, but affects status flags. On the other hand, \texttt{lea~reg,~[reg+0]} has the same latency, but does not have any such side-effects. For \texttt{class2}, it is more difficult to find a dummy instruction, as there is no no-op instruction with similar latency to \texttt{div}, which affects status flags. As a workaround, \texttt{div} instructions are enclosed with infrastructure that ensures that flags are correctly preserved, making \texttt{class2} instructions and their dummy equivalents take two instruction slots.

\section{Evaluation}
\label{sec:evaluation}

We now evaluate the security and performance of \tool{}. We use the same processors as in our latency analysis in \Cref{sec:uniform-code-blocks:instruction-analysis}.

\begingroup
\begin{table*}[ht!]

    \caption{Benchmark of several example programs with the different \tool{} variants on our AMD machine. All times are given in microseconds, the overheads are factors relative to the original execution time (orig).}
    \label{tab:performance}
    \centering
    \small
    
    \begin{tabular}{l r r >{\bfseries}r r >{\bfseries}r r >{\bfseries}r r >{\bfseries}r r >{\bfseries}r} % l*{11}{r}
    \toprule
    
    \multirow{2}[3]{*}{\hspace*{-0.5em}Target} & \multirow{2}[3]{*}{\hspace*{-0.5em}\timebase{}} & \multicolumn{2}{c}{\toolIshort{}} & \multicolumn{2}{c}{\toolIIshort{}} & \multicolumn{2}{c}{\toolIIIshort{}} & \multicolumn{2}{c}{\toolIVshort{}} & \multicolumn{2}{c}{\toolVshort{}}\\
    \cmidrule(lr){3-4}
    \cmidrule(lr){5-6}
    \cmidrule(lr){7-8}
    \cmidrule(lr){9-10}
    \cmidrule(lr){11-12}
      &  & \timetool{} & \normalfont{\overhead{}} & \timetool{} & \normalfont{\overhead{}} & \timetool{} & \normalfont{\overhead{}} & \timetool{} & \normalfont{\overhead{}} & \timetool{} & \normalfont{\overhead{}} \\
      
    \midrule 
    \midrule
    
    \multicolumn{12}{l}{\hspace*{-0.5em}\underline{\texttt{small}}}\\
    \texttt{matmul} & 0.053 & 37 & 703x & 38 & 720x & 38 & 706x & 35 & 660x & 38 & 708x \\
    \texttt{modexp} & 0.217 & 49 & 229x & 50 & 231x & 37 & 169x & 35 & 163x & 37 & 171x \\
    
    \midrule
    
    \multicolumn{12}{l}{\hspace*{-0.5em}\underline{\texttt{mbedTLS}}}\\
    \texttt{aes}        & 0.134 & 401 & \num{2983}x & 403 & \num{3010}x & 196 & \num{1463}x & 192 & \num{1424}x & 217 & \num{1607}x \\
    \texttt{base64}     & 0.785 & 363 & 454x & 440 & 559x & 413 & 526x & 397 & 506x & 425 & 541x \\
    \texttt{cc20}       & 0.455 & \num{2045} & \num{4526}x & \num{2068} & \num{4806}x & \num{1802} & \num{4208}x & \num{1799} & \num{4205}x & \num{1869} & \num{4403}x \\
    \texttt{ecdh}       & 890.6 & 90M & \num{101019}x & 90M & \num{102910}x & 68M & \num{77361}x & 69M & \num{78439}x & 78M & \num{88852}x \\
    \texttt{rsa}        & \num{1061.9} & 50M & \num{47107}x & 59M & \num{55959}x & 48M & \num{45716}x & 48M & \num{44694}x & 83M & \num{78750}x \\
    
    \bottomrule
    \end{tabular}
    
\end{table*}
\endgroup

\subsection{\tool{} Variants}
\label{sec:evaluation:variants}
In order to understand the characteristics of different security levels, we devised five variants of \tool{}, which can be selected when compiling the target program. The first variant is named \toolI{}, where we build code blocks greedily by putting a memory read and a write at the beginning and then adding as many instructions until the code block is filled.
This is the Obfuscuro approach. %
The second variant \toolII{} is similar, but uses a fixed instruction count instead of filling the code block, to thwart instruction counting attacks.
We set the instruction count to 10 for a code block size of 160 bytes (20 8-byte slots).

The variant \toolIII{} is the first to actually enforce the pattern generated by our algorithm  in \Cref{sec:uniform-code-blocks:pattern-generation}, and thus is resistant to Nemesis-style~\cite{DBLP:conf/ccs/BulckPS18} latency measurement attacks.
Variant \toolIV{} adds no-op padding to ensure that all instructions are aligned at 8-byte boundaries, to eliminate the risk of alignment-based timing differences~\cite{DBLP:conf/uss/PudduSHC21}. \toolV{} uses the same code blocks, but additionally enforces ciphertext freshness.

\subsection{Security Evaluation}
\label{sec:evaluation:security}

We built a small example program, called \texttt{modexp}, which computes a modular exponentiation of integers with a standard leaky square \& multiply algorithm. The given implementation has a secret-dependent branch, either performing a ``square, mod'' operation, or ``square, multiply, mod''. For different \tool{} variants, we analyze how an attacker can try to distinguish the resulting code blocks after \num{10000} executions. Given the constant-time ORAM implementation, they are not able to use timing attacks to find out which block is currently executed, and the ciphertext side-channel countermeasures prevent straightforward labeling. Thus, the attacker may resort to single-stepping, to count instructions or tell them apart by measuring their latency differences. \Cref{fig:block-comparison} shows the result for the aforementioned operations.
We did not observe penalties from instruction alignment~\cite{DBLP:conf/uss/PudduSHC21} in our examples, so we left out the measurements for \toolIII{}, which look identical to \toolIV{}.

We see that for variant \toolI{}, which builds blocks greedily until they are filled (the Obfuscuro approach), the different instruction counts are immediately apparent to an attacker, which confirms our findings in \Cref{sec:defenses:obfuscuro-attack}.
Variant \toolII{} addresses this by using a fixed instruction count per block. However, the \texttt{div} is placed at another offset within the block, allowing the attacker to distinguish them via a latency measurement.
Finally, with \toolIV{}, instructions are assigned to fixed slots with identical alignment, and the $t$-test reports no leakage for \texttt{div}. Due to noise, we observe false-positive differences between \texttt{xor} and \texttt{imul}, which disappear when conducting more measurements.

We thus conclude that, within the limits discussed in \Cref{sec:uniform-code-blocks:instruction-analysis}, an attacker is not able to distinguish code blocks which use a fixed and aligned code block pattern, making \tool{} secure in our attacker model.

\subsection{Performance}
\label{sec:evaluation:performance}

To analyze the performance impact of \tool{}, we applied it to a number of targets. We first evaluated two small example programs, \texttt{modexp} and \texttt{matmul} from Obfuscuro. To show that \tool{} works with a large real-world library, we applied it to a representative set of cryptographic primitives from MbedTLS~\cite{target-mbedtls}.
We ran \num{1000} measurements per target and computed the mean execution times and overheads. The results are summarized in \Cref{tab:performance}. Additionally, \Cref{tab:other-stats} shows the compilation time, code size and memory usage overhead of (the most expensive) variant \toolV{}.

As is visible in the results, performance overhead is rather high, especially for large programs.
We found that the main contributors to the overhead are the ORAM queries and machine clears due to self-modifying code detection (see \Cref{sec:discussion:smc}).
Clearly, a large program which handles lots of data also leads to more expensive ORAM queries per code block and per memory access, and thus a higher overhead.
We also note that \tool{} variants with a fixed block pattern often perform better than the Obfuscuro-equivalent baseline \toolI{}. This is mostly due to memory accesses placed at more convenient locations, leading to comparatively fewer code blocks.

As ORAM queries are a main bottleneck, finding a more efficient ORAM implementation is highly desirable.
For example, it was shown that ORAM queries can be sped up notably by offloading them to an FPGA~\cite{DBLP:conf/ccs/OhAPLP20}. We leave development of faster side-channel resistant ORAM algorithms to future work.

The compilation time directly depends on the parameters used for the genetic algorithm and the number of call trees which need an individual optimal code block pattern. While we used a rather high threshold of 10 seconds for the pattern search, we found that it can be safely reduced to 2 seconds without any loss of quality for the non-asymmetric examples, reducing the compilation overhead by 80\%

We conclude that the performance overhead may be too high for large and complex programs such as asymmetric cryptography. However, for medium-sized programs which are only executed on occasion or run asynchronously (such as a license check running on an end user's system), the overhead can be justified given the broad security guarantees of \tool{}. %
By underlining the high usefulness of ORAM for side-channel protection, we hope to inspire further research to identify efficient ORAM schemes.

\section{Discussion}
\label{sec:discussion}

\begin{table}[t]
    \caption{Build time, code size and memory usage of the example programs protected with variant \toolV{}. Build time and code size are given for the entire library binary, while memory usage is estimated for the isolated primitives.}
    \label{tab:other-stats}
    \centering
    \small
    
    \begin{tabular}{ l r r r r } 
    \toprule
    
    \multirow{2}[3]{*}{\hspace*{-0.5em}Target} & \multirow{2}[3]{*}{\hspace*{-1em}\shortstack{Build time\\ (s)}} & \multirow{2}[3]{*}{\hspace*{-0.5em}\shortstack{Code size\\ (KB)}} & \multicolumn{2}{c}{Memory usage (KB)} \\
    \cmidrule(lr){4-5}
      &  &  & original & instrum. \\
      
    \midrule 
    \midrule
    
    \multicolumn{1}{l}{\hspace*{-0.5em}\underline{\texttt{small}}} & 0.1 & 6.4   & &\\
    \texttt{matmul} & 2.5 & 11.1  & 2204 & 2216 \\
    \texttt{modexp} & 2.2 & 12.7  & 2204 & 2240 \\
    
    \midrule

    \multicolumn{1}{l}{\hspace*{-0.5em}\underline{\texttt{mbedTLS}}}    & 5.9 & 697 &  &  \\
    \texttt{aes}        & 31.5 & 769 & 2472 & 2712 \\
    \texttt{base64}     & 31.5 & 732 & 2392 & 2504 \\
    \texttt{cc20}       & 28.1 & 864 & 2528 & 2768 \\
    \texttt{ecdh}       & 29.0 & 1915 & 2752 & 5916 \\
    \texttt{rsa}        & 53.5 & 2818 & 2788 & 6384 \\
    
    \bottomrule
    \end{tabular}
\end{table}

\subsection{Circumventing Self-Modifying Code Detection}
\label{sec:discussion:smc}
Both Intel and AMD CPUs guarantee that self-modifying code (SMC) is executed correctly, i.e., the CPU always runs the latest version of the machine code that exists in memory. To achieve this, the CPU vendors deploy various mechanisms, e.g., checking whether pending stores touch addresses currently present in the pipeline (AMD) or in the instruction cache (Intel).
These are barely documented and cause severe penalties, as upon detecting SMC the entire CPU pipeline gets flushed (machine clear)~\cite{DBLP:conf/uss/RagabBBG21}.

As we overwrite the code scratchpad with the next code block, we trigger an SMC condition. Memory fences and serializing instructions (as recommended by the documentation) after the code scratchpad store do not fix this issue, which we suspect is due to out-of-order execution and possibly prefetching.
We found that the code scratchpad rotation from our ciphertext side-channel protection significantly improves performance.
On AMD, we could further reduce the observed penalties by adding a number of dummy \texttt{div} instructions before jumping into the new code block, to prevent the CPU from executing it out-of-order before the store has completed. Still, on AMD, SMC machine clears are responsible for up to 50\%
We leave a thorough analysis of the SMC detection mechanisms and suitable workarounds to future work. Circumventing the performance penalties of SMC detection is an interesting research question, the answer to which may help speed up both \tool{}-like code hardening frameworks and common just-in-time compilers.

\subsection{Integrity Protection}
Another attack class relevant for TEEs are \emph{fault injection attacks} which try to corrupt the data, code or computations within the enclave. While we did not include them in our proof-of-concept implementation of \tool{}, its modular structure and the clear separation of concerns is particularly suited for implementing effective countermeasures.

Attacks against memory integrity such as Rowhammer~\cite{DBLP:conf/isca/KimDKFLLWLM14} bypass the CPU protections and flip bits directly in physical memory. While Intel SGX checks memory integrity, AMD SEV fully relies on its memory encryption to prevent targeted modifications. However, even if the attacker is not able to place their own plaintext, they can still tamper with data in order to break cryptographic implementations such as RSA-CRT~\cite{DBLP:conf/wistp/KimQ07}. %
\tool{} can address this by introducing an own layer of integrity checks in the ORAM controllers and the code blocks.

Preventing undervolting attacks like Plundervolt~\cite{DBLP:conf/sp/MurdockOGBGP20} which target the integrity of computations in the processor itself is more involved, but possible.
For example, in the Plundervolt paper the authors stated that they could fault multiplication instructions, but not simpler arithmetic like addition or shifts.
To harden \tool{} against such attacks, one could systematically identify such vulnerable instructions, and replace them in the controllers and code blocks by sequences of instructions which are less susceptible to faulting. 
In combination with a memory integrity protection as discussed above and the built-in control flow obliviousness which complicates targeting vulnerable code sections, this should greatly reduce the attack surface.

\subsection{Transient Execution Attacks}
While we considered transient execution attacks such as Spectre~\cite{DBLP:conf/sp/KocherHFGGHHLM019,DBLP:conf/uss/CanellaB0LBOPEG19} out-of-scope, \tool{} already provides good protection against such attacks: Due to partitioning the code into branch-free blocks, which are pulled from an ORAM in a rather expensive operation, the attacker has little opportunity to trigger speculative execution of code blocks. The indirect branches necessary to jump between the code block and controller are vulnerable to Spectre-BTB, but this can be addressed by \texttt{endbr64} instructions at the controller entry points, the code block entry point and in the code block memory accesses. \texttt{endbr64} was introduced with Intel CET and enforces that indirect jumps always end at such an instruction, greatly reducing the number of gadgets reachable by manipulating pointers of indirect jumps (even speculatively~\cite[17.3.8]{intel2023sdm}).
This leaves hardening the controller, which can be done using conventional methods.

\section{Related Work} %
\label{sec:related-work}
A number of publications propose \emph{single-stepping countermeasures} to thwart fine-grained measurements of enclaves and their state.
Proposals like T-SGX~\cite{DBLP:conf/ndss/Shih0KP17} or MoLE~\cite{DBLP:conf/acsac/Lang0MLWL22} try to protect leaking executions from being interrupted, but they depend on TSX, which is known to introduce vulnerabilities and was therefore disabled on lots of processors via a microcode update~\cite{intel2023tsx}.
Varys~\cite{DBLP:conf/usenix/OleksenkoTKSF18}, Déjà-Vu~\cite{DBLP:conf/ccs/ChenZRZ17} and HyperRace~\cite{DBLP:conf/sp/ChenWCCZWLL18} check AEX to prevent single-stepping, but with a limited scope.
AEX-Notify~\cite{DBLP:conf/uss/ConstableBCXXAK23,intel-aex-notify} successfully deployed a hardware-assisted way of making enclaves interrupt-aware to mitigate single-stepping attacks. However, it is limited to SGX.
For Intel TDX, the TDX module ensures that single-stepping is no longer precise and deterministic~\cite{intel-tdx-manual} by allowing to execute a random number of instructions after an interrupt, but this only reduces the accuracy of stepping and does not prevent side-channel leakage in general.
Nothing similar to the TDX module has yet been proposed for AMD SEV or ARM CCA. %
Therefore, other ways to protect the code running in TEEs are needed to ensure that secrets are not leaked during execution.

\emph{Compiler-based side-channel mitigations} have been proposed for various protection levels.
SGX-Shield~\cite{DBLP:conf/ndss/SeoLKSSHK17}, Klotski~\cite{DBLP:conf/asplos/ZhangSYZS020} and deterministic multiplexing~\cite{DBLP:conf/ccs/ShindeCNS16} only hide accesses to code at page level granularity and thereby do not protect against cache attacks.
Tools that result in data-oblivious execution include Constantine~\cite{DBLP:conf/ccs/BorrelloDQG21} as well as SGX-specific approaches like Obliviate~\cite{DBLP:conf/ndss/AhmadKSL18}, Raccoon~\cite{DBLP:conf/uss/RaneLT15}, ZeroTrace~\cite{DBLP:conf/ndss/SasyGF18} and DR.SGX~\cite{DBLP:conf/acsac/BrasserCDFKS19}.
However, they do not aim to protect the executed code itself, so attackers can still infer information about the used algorithms.
As discussed in \Cref{sec:defenses:obfuscuro-attack}, Obfuscuro~\cite{DBLP:conf/ndss/AhmadJXZSL19} has the goal of code and data obliviousness, but does neither protect against single-stepping attacks nor ciphertext side-channel leakage.
A Nemesis countermeasure for specific embedded targets has been proposed by Winderix et al.~\cite{DBLP:conf/eurosp/WinderixMP21}, who equalize secret-dependent branches by aligning basic blocks in a way such that their latency profiles become indistinguishable.
This has also been adapted in a thorough analysis and mitigation approach~\cite{DBLP:conf/eurosp/BognarWBP23}.
However, their work is tailored to embedded targets without critical optimizations such as caches or out-of-order execution, as they require deterministic instruction timings. On more complex processors, an attacker can mount a cache attack to observe which part of a balanced branch is executed or which memory address is accessed, bypassing the latency trace protection.

\section{Conclusion}
\label{sec:conclusion}

In this work, we have showcased \tool{}, a compiler-based drop-in software-level defense against various side-channel based attacks.
Our approach is based on oblivious code execution and data accesses by using Linear ORAM together with uniform code blocks which are indistinguishable even for strong side-channel attackers in a TEE scenario.
We have shown that \tool{} successfully safeguards implementations against different attack classes.
Due to its modular structure, \tool{} can be easily extended with defenses for other attack classes in the future.
In summary, even for application developers without expertise in side-channel defense, \tool{} provides a way to automatically protect implementations against all relevant attack classes.

\ifthenelse{\equal{\isAnon}{1}}{
	
}{
	\ifUsenix    
        \section*{Acknowledgements}
    \else 
        \ifCLASSOPTIONcompsoc
            \section*{Acknowledgments}
        \else
            \section*{Acknowledgment}   
        \fi
    \fi
  
  We would like to thank the anonymous reviewers and our shepherd for their valuable feedback.
  This work has been supported by Deutsche Forschungsgemeinschaft (DFG) under grants 427774779 and 439797619, and by Bundesministerium für Bildung und Forschung (BMBF) through the ENCOPIA and SAM-Smart projects.
} %

\ifUsenix    
    \bibliographystyle{plain}{%
	\footnotesize%
	\bibliography{ref}
    }
\else 
    \bibliographystyle{IEEEtran}{%
	\footnotesize%
	\bibliography{ref}%
    }
\fi

\appendices

\newpage
\section{SEV-Step Measurements}

\begin{figure}[h!]
    \centering
    \begin{subfigure}[]{0.45\textwidth}
        \centering
        \includegraphics[width=\textwidth]{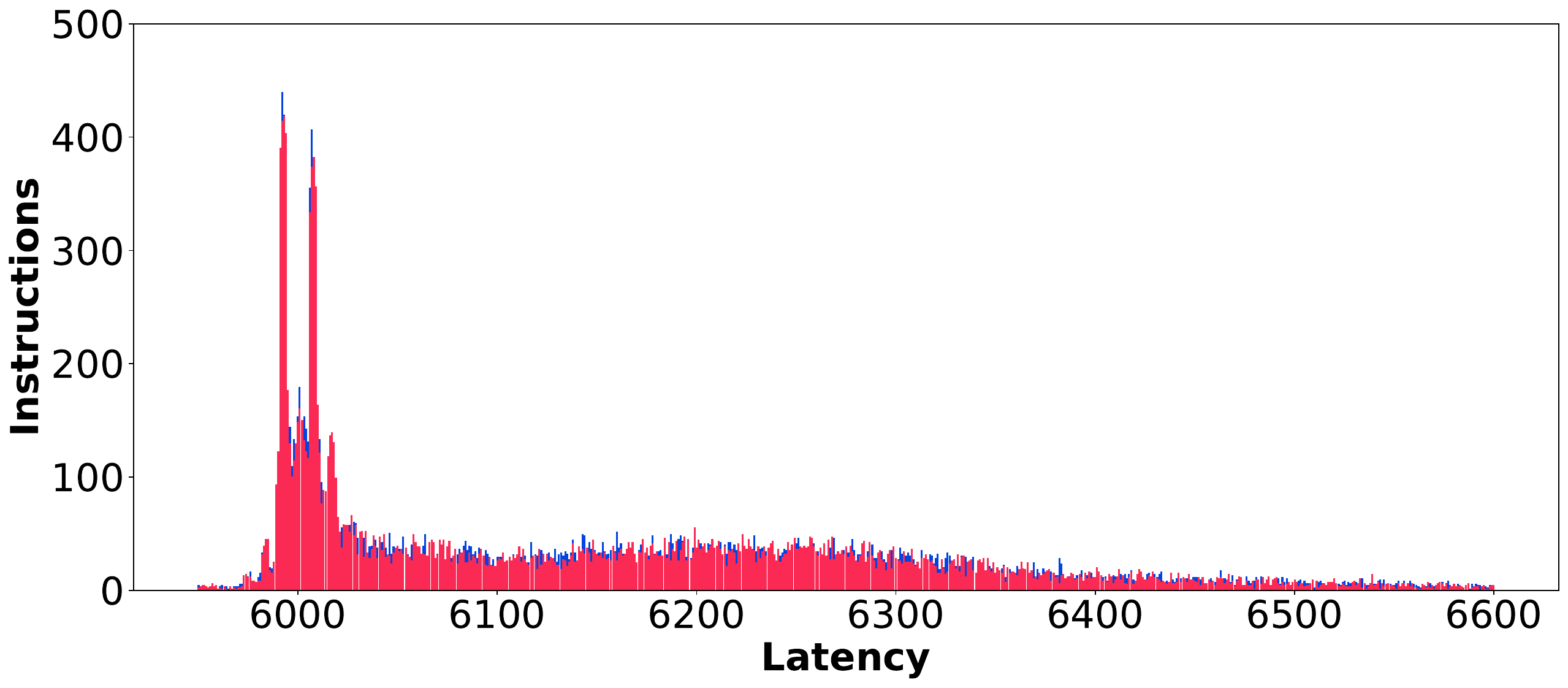}
        \caption{Histogram of \texttt{mov} vs. \texttt{imul}}
        \label{fig:sev:mov-imul}
    \end{subfigure}
    \\
    \begin{subfigure}[]{0.45\textwidth}
        \centering
        \includegraphics[width=\textwidth]{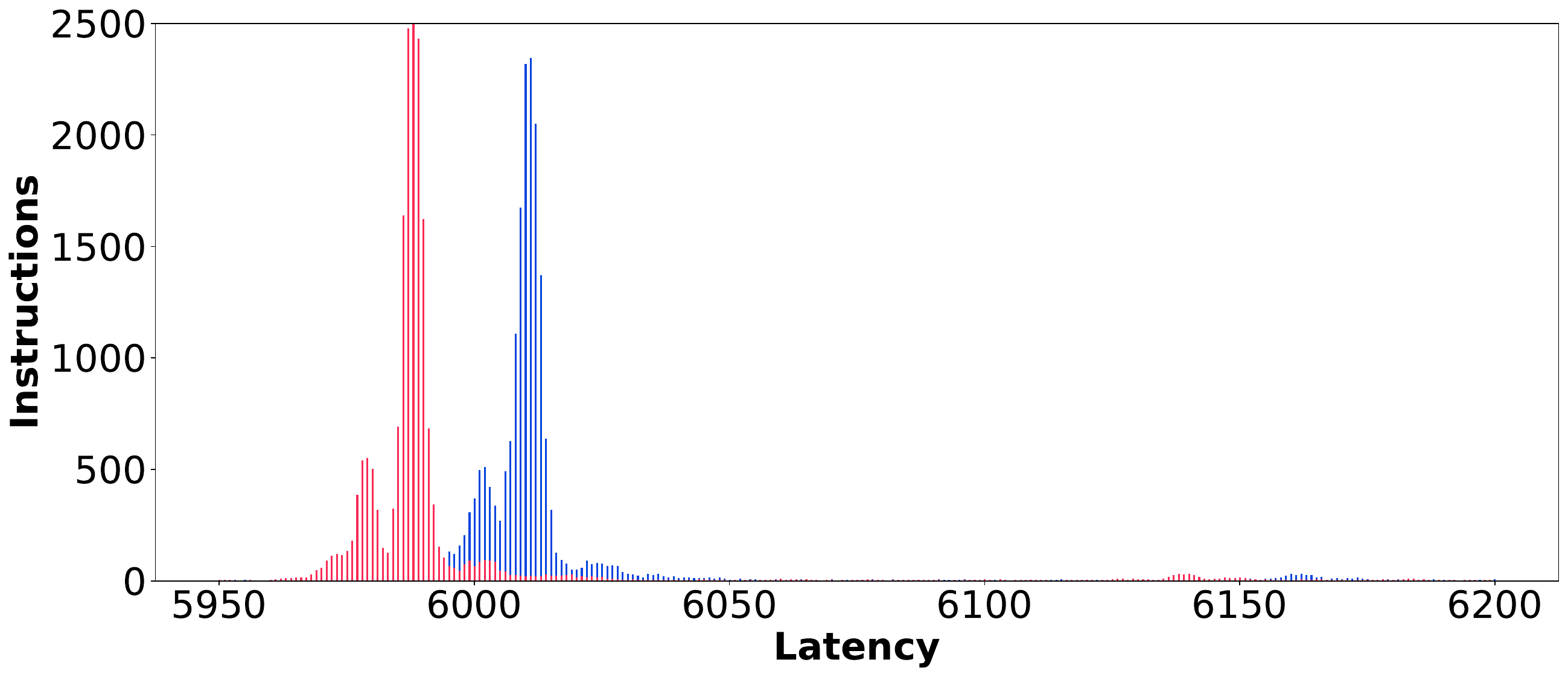}
        \caption{Histogram of \texttt{mov} vs. \texttt{div}}
        \label{fig:sev:mov-div}
    \end{subfigure}
    \\
    \begin{subfigure}[]{0.45\textwidth}
        \centering
        \includegraphics[width=\textwidth]{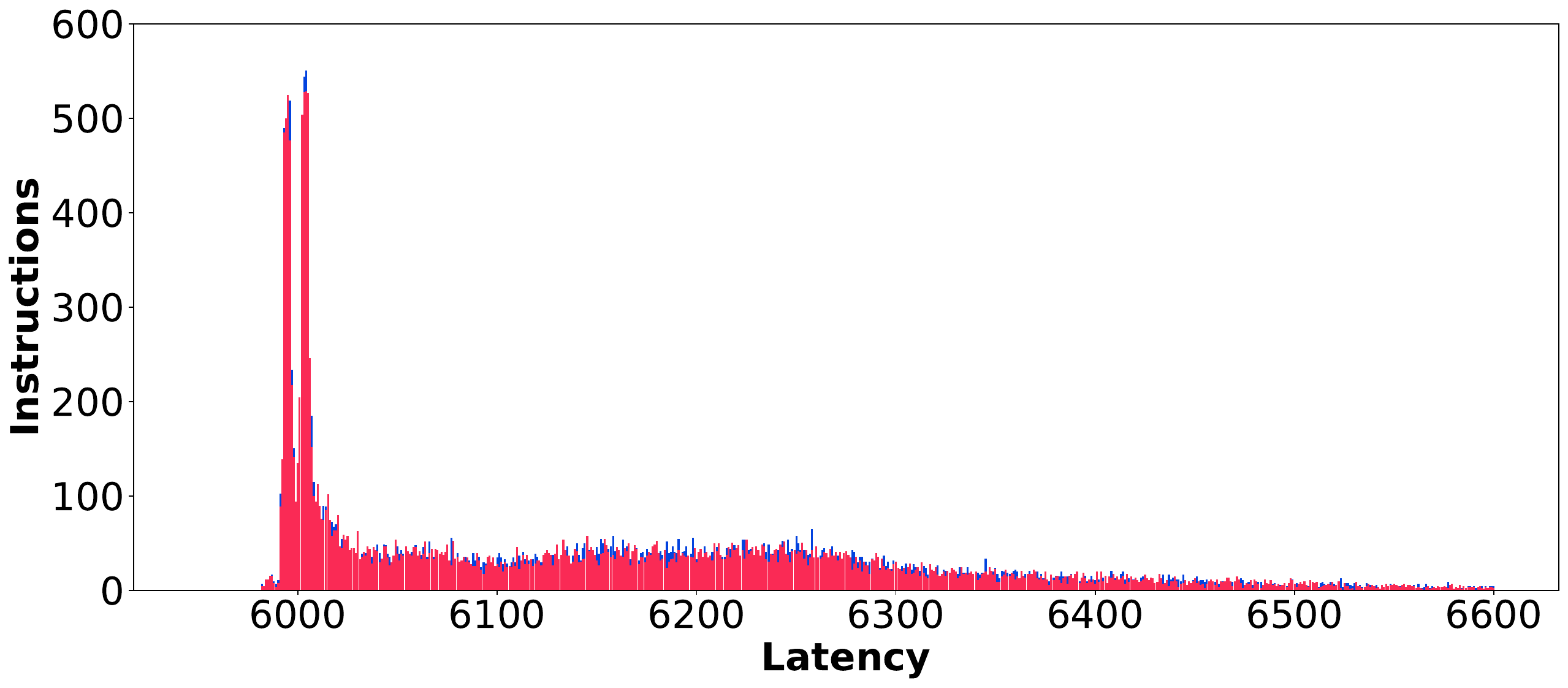}
        \caption{Histogram of \texttt{inc} vs. \texttt{inc}}
        \label{fig:sev:inc-inc}
    \end{subfigure}
    
    \caption{Histograms for our instruction latency experiments on AMD SEV, analogous to those depicted in \Cref{fig:instruction-analysis} for Intel SGX.}
    \label{fig:sev-measurements}
\end{figure}

\newpage

\section{Meta-Review}

The following meta-review was prepared by the program committee for the 2024 IEEE Symposium on Security and Privacy (S\&P) as part of the review process as detailed in the call for papers.

\subsection{Summary}
This paper presents Obelix, a framework to mitigate side-channel attacks in trusted execution environments (TEEs). By leveraging a linear oblivious RAM model and enforcing uniform code blocks, Obelix can prevent attackers from gaining insights into both executed code and accessed data.

\subsection{Scientific Contributions}
\begin{itemize}
    \item Addresses a Long-Known Issue
    \item Provides a Valuable Step Forward in an Established Field
    \item Creates a New Tool to Enable Future Science
\end{itemize}

\subsection{Reasons for Acceptance}
\begin{enumerate}
    \item This paper addresses a long-known issue. Recent work has revealed numerous vulnerabilities that attackers can exploit to tamper with the executed code or secret data in trusted execution environments. This paper aims to protect both the code and data against relevant side-channel vulnerabilities in TEEs.
    \item The paper provides a valuable step forward in an established field. There are numerous existing countermeasures, but most of them focus on protecting the secret data, and only a few are deployed into the hardware by vendors. Obelix takes a holistic view of the vulnerabilities and designs solutions that can protect both code and data against a wide range of TEE attacks at the software level.
    \item The paper creates a new tool to enable future science. Obelix is implemented as an LLVM compiler extension, and the authors will make it publicly available, thereby facilitating future research.
\end{enumerate}

\subsection{Noteworthy Concerns}
While Obelix offers valuable security enhancements, its deployment could lead to considerable overhead. This work evaluates that with micro-benchmarks, but it lacks thorough evaluations of real-world, full-fledged applications.

\end{document}